\documentclass[aps,preprint]{revtex4}%
\usepackage{amsfonts}%
\usepackage{amsmath}%
\usepackage{amssymb}%
\usepackage{graphicx}

\begin{document}

\input epsf.tex    

\input psfig.sty

\title{Physics of Ultra-Peripheral Nuclear Collisions}

\markboth{Carlos A. Bertulani, Spencer R. Klein and Joakim
Nystrand}{Physics of Ultra-Peripheral Nuclear Collisions}
\author{\tt Carlos A.
Bertulani$^{1}$, Spencer R. Klein$^{2}$
 and Joakim Nystrand$^{3}$}
\email{bertulani@physics.arizona.edu,srklein@lbl.gov,joakim.nystrand@ift.uib.no}
\affiliation{$^{1}$ Department of Physics,
University of Arizona, 1118 E. 4th St., Tucson, AZ 85721, USA\\
$^{2}$ Nuclear Science Division, LBNL, 1 Cyclotron Road,
Berkeley CA 94720, USA\\
$^{3}$ Department of Physics and Technology, University of Bergen,
Allegaten 55, 5007 Bergen, Norway}

\begin{abstract}

Moving highly-charged ions carry strong electromagnetic fields that
act as a field of photons.  In collisions at large impact
parameters, hadronic interactions are not possible, and the ions
interact through photon-ion and photon-photon collisions known as
{\it ultra-peripheral collisions} (UPC).  Hadron colliders like the
Relativistic Heavy Ion Collider (RHIC), the Tevatron and the Large
Hadron Collider (LHC) produce photonuclear and two-photon
interactions at luminosities and energies beyond that accessible
elsewhere; the LHC will reach a $\gamma p$ energy ten times that of
the Hadron-Electron Ring Accelerator (HERA).  Reactions as diverse
as the production of anti-hydrogen, photoproduction of the $\rho^0$,
transmutation of lead into bismuth and excitation of collective
nuclear resonances have already been studied.  At the LHC, UPCs can
study many types of `new physics.'

\end{abstract}

\maketitle


\section{Introduction}

In 1924, Enrico Fermi, then 23 years old, submitted a paper ``On the Theory of
Collisions Between Atoms and Elastically Charged Particles" to Zeitschrift
f\"ur Physik \cite{Fe24}. This paper does not appear in his ``Collected
Works". Nevertheless, it is said that this was one of Fermi's favorite ideas
and that he often used it later in life \cite{MW03}. In this publication,
Fermi devised a method known as the \textit{equivalent (or virtual) photon
method}\textrm{, where he treated the electromagnetic fields of a charged
particle as a flux of virtual photons. Ten years later, Weiszs\"acker and
Williams extended this approach to include ultra-relativistic particles, and
the method is often known as the \textit{Weizs\"acker-Williams method}
\cite{WW34}. }

\textrm{A fast-moving charged particle has electric field vectors pointing
radially outward and magnetic fields circling it. The field at a point some
distance away from the trajectory of the particle resembles that of a real
photon. Thus, Fermi replaced the electromagnetic fields from a fast particle
with an \textit{equivalent flux of photons}. The number of photons with energy
$\omega$, $n(\omega)$, is given by the Fourier transform of the time-dependent
electromagnetic field. The virtual photon approach used in quantum
electrodynamics (QED) to describe, e.g. atomic ionization or nuclear
excitation by a charged particle can be simply described using Fermi's
approach. }

\textrm{When two nuclei collide, two types of electromagnetic processes can
occur. A photon from one ion can strike the other, Figure \ref{diags}(a), or,
photons from each nucleus can collide, in a photon-photon collision, as in
Figure \ref{diags}(b). }

\textrm{Ultra-peripheral hadron-hadron collisions will provide unique
opportunities for studying electromagnetic processes. At the LHC,
photon-proton collisions will occur at center of mass energies an order of
magnitude higher than are available at existing accelerators, and photon-heavy
ion collisions will reach 30 times the energies available at fixed target
accelerators. The electromagnetic fields of heavy-ions are very strong, so
reactions involving multi-photon excitations can be studied. }

\textrm{Ultra-relativistic heavy-ion interactions have been used to study
nuclear photoexcitation (e.g. to a Giant Dipole Resonance), and
photoproduction of hadrons. Coulomb excitation is a traditional tool in low
energy nuclear physics. The strong electromagnetic fields from a heavy ion
allow for the study of multi-photon excitation of nuclear targets. This allows
the study of high-lying states in nuclei, e.g. the double-giant resonance
\cite{BB86,sch93}. Multiple, independent interactions among a single ion pair
are also possible. Reactions like multiple vector meson production can be used
for studies involving polarized photons. The high photon energies can be used
to study the gluon density in heavy nuclei \cite{GoBe02} at low Feynman-$x$. }

\textrm{The cross section for photoproduction is
\begin{equation}
\sigma_{X}=\int d\omega\frac{n\left( \omega\right)  }{\omega}\sigma
_{X}^{\gamma}\left( \omega\right)  \;,\label{eq:photonuclear}%
\end{equation}
where $\sigma_{X}^{\gamma}\left( \omega\right)  $ is the photonuclear cross
section. }

\textrm{\begin{figure}[t]
\textrm{\centerline{\psfig{figure=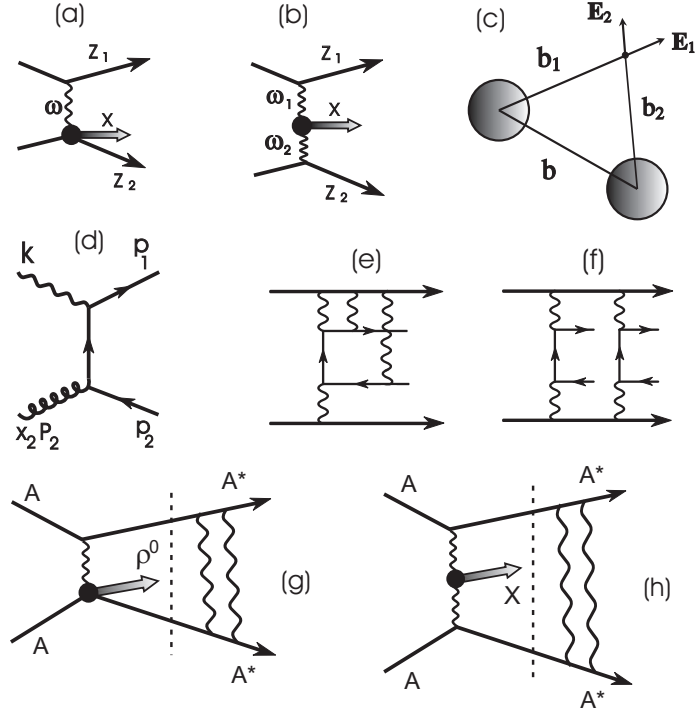,height=3.7 in}} }%
\caption{{\small (a) One-photon and (b) two-photon processes in heavy ion
collisions. (c) Geometrical representation of the photon fluxes at a point
outside nuclei 1 and 2, in a collision with impact parameter $b$. The electric
field of the photons at that point are also shown. (d) Feynman diagram for $q
\overline{q}$ production through photon-gluon fusion to leading order. (e,f)
Example of higher order corrections to pair-production: (e) Coulomb
distortion, and (f) production of multiple pairs. (g) The dominant diagram for
$Au + Au \rightarrow Au^{*} + Au^{*} + \rho^{0}$ and (h) for $Au + Au
\rightarrow Au^{*} + Au^{*} + e^{+}e^{-}$ or a meson $X$. The dotted lines in
panels (g) and (h) show how the mutual Coulomb nuclear excitation factorizes
from the particle production.}}%
\label{diags}%
\end{figure}}

\textrm{Photon-photon (or ``two-photon'') processes have long been studied at
$e^{+}e^{-}$ colliders. They are an excellent tool for many aspects of meson
spectroscopy and tests of QED. At hadron colliders, they are also used to
study atomic physics processes, often involving electrodynamics in strong
fields. One striking success was the production of antihydrogen atoms at
CERN's\footnote{\textrm{European Organization for Nuclear Research (Conseil
Europ\'{e}enne pour la Recherche Nucl\'{e}aire.)}} LEAR\footnote{\textrm{Low
Energy Antiproton Ring}} \cite{Bau96} and at the Fermilab Tevatron
\cite{Blan98}. At the highest energy colliders, reactions like $\gamma
\gamma\rightarrow X$ may be used to probe the quark content and spin structure
of mesons resonances. Production of meson or baryon pairs can also probe the
internal structure of hadrons. At the LHC, electroweak processes such as
$\gamma\gamma\rightarrow W^{+}W^{-}$ may be probed. The cross section for
two-photon processes is \cite{BB88}
\begin{equation}
\sigma_{X}=\int d\omega_{1} d\omega_{2} \frac{n\left( \omega_{1}\right)
}{\omega_{1}} \frac{n\left( \omega_{2}\right)  }{\omega_{2}} \sigma
_{X}^{\gamma\gamma}\left( \omega_{1},\omega_{2}\right)
\;,\label{eq:two-photon}%
\end{equation}
where $\sigma_{X}^{\gamma\gamma}\left( \omega_{1},\omega_{2}\right)  $ is the
two-photon cross section. }

\textrm{Fermi's method has found application beyond the realms of QED. It has
been extended to strong interactions between nuclei in peripheral collisions.
These interactions are mainly mediated by pion exchange, and an
\textit{equivalent pion method} has been applied, to describe subthreshold
pion production in nucleus-nucleus collisions \cite{Pir80}. Feshbach used the
term \textit{nuclear Weiszs\"acker-Williams method} to describe excitation
processes induced by the nuclear interaction in peripheral collisions of heavy
ions \cite{Fesh77}. More recently, a \textit{non-Abelian
Weiszs\"acker-Williams field} was used to describe the boosted gluon
distribution functions in nuclear collisions \cite{MV94}. }

\textrm{Since Fermi's original work, much progress has been achieved in this
field, especially with the advent of relativistic heavy ion accelerators like
the Bevalac accelerator at Lawrence Berkeley National Laboratory (LBNL).
Intermediate energy processes have been explored at heavy ion accelerators at
NSCL/MSU, GANIL, RIKEN, and GSI\footnote{\textrm{NSCL/MSU: National Science
Cyclotron Laboratory at Michigan State University, GANIL: Grand Accelerateur
National d'Ions Lourds in Caen/France, RIKEN: The Institute of Physical and
Chemical Research, Wako, Saitama/Japan, GSI: Gesellschaft fuer
Schwerionenforschung, Darmstadt/Germany}}. These facilities have explored the
collective excitation and electromagnetic fragmentation of nuclei, and studied
many reactions that occur in the sun, supernovae, and the big bang.
Experimental studies of higher-energy processes have recently begun at
Brookhaven's RHIC. These studies have included vector meson spectroscopy and
production of $e^{+}e^{-}$ pairs. In the next few years, CERNs LHC will begin
operations, allowing for the study of heavy mesons, measurements of gluon
distributions in nuclei, and searches for a host of `new physics' processes. }

\textrm{This review will discuss these experiments, their theoretical
interpretation, and some future possibilities in this field. UPCs have been
previously reviewed by a number of authors \cite{BB88,
KGS97,BHT98,BHTSK02,Hot02}. }

\subsection{\textrm{The Photon Flux}}

\textrm{The flux of equivalent photons from a charged particle is determined
from the Fourier transform of the electromagnetic field of the moving charge.
The fields of a relativistic particle Lorentz contract toward a co-moving
pancake, as is shown in Fig. \ref{fig:pancake}. The photon energy spectrum
depends on the time a target particle spends in this pancake, \textit{i.e.} on
the minimum distance between the target and the charge and on the projectile
velocity; the minimum photon wavelength is the width of the pancake at the
target. At an ion-ion separation (impact parameter) $b$, the interaction time
is $\Delta t \sim b/(\gamma v)$. In the lab frame, the maximum photon energy
is
\begin{equation}
\omega^{\max}=\frac{\hbar}{\Delta t} \sim\frac{\gamma\hbar v}{b},
\end{equation}
where $\gamma$ is the Lorentz factor of the particle, $\gamma= (1-v^{2}%
/c^{2})^{-1/2}$. In the target frame, this equation applies, as long as
$\gamma$ is taken as the boost to go from the frame of one nucleus to the
other ($\gamma=2\gamma_{collider}^{2}-1$). }

\textrm{\begin{figure}[t]
\textrm{\centerline{\psfig{figure=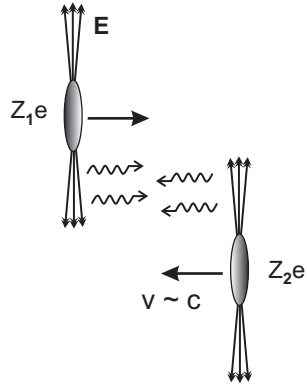,height=2.0 in}} }\caption{Highly
energetic charged particles have Lorentz contracted electric fields. The
interaction of these fields can be replaced by the interaction of real (or
quasi-real) photons.}%
\label{fig:pancake}%
\end{figure}}

\textrm{For a grazing collision, where the two nuclei barely touch, we can
take $b_{min}=2R_{A}$, and the maximum photon energy is $\gamma\hbar v/2R_{A}$
($R_{A}=$ nucleus radius). The maximum photon energy is about $\hbar/(2R_{A} A
m_{p}c)$ of the ion energy. Here, $Am_{p}$ is the ion mass. For heavy-ions,
$R_{A}\approx7$ fm so $\omega^{max}$ is about $0.03/A$ of the ion energy. For
protons, $R_{A}$ is not well defined, but taking $\omega^{max}$ to be 10\% of
the proton energy is a reasonable rule-of-thumb. RHIC (see Table 1) can reach
photon-gold center of mass energies up to about 30 GeV per nucleon, and
photon-proton center of mass energies up to 300 GeV. These energies are
slightly higher than are available at fixed target accelerators and at HERA
respectively. At the LHC, the corresponding energies are 1 TeV and 10 TeV
respectively, more than an order of magnitude higher than is available
elsewhere. 0 \begin{table}[ptb]
\caption{Some ion species, maximum energy and luminosity for several
accelerators \cite{PDG04}. Also shown are the maximum effective $\gamma$p and
$\gamma\gamma$ energies. For proton beams, the maximum effective photon energy
is taken to be 10\% of the proton energy, although there is some flux at
higher energies. The CERN SPS is a fixed target accelerator; the effective
luminosity depends on the target thickness. Not mentioned here are
lower-energy accelerators, where photon exchange processes have also been
studied.}%
\label{accelerators}
\begin{center}
\textrm{%
\begin{tabular}
[c]{rccccc}\hline\hline
Accelerator & Ions & Max. Energy (CM) & Luminosity & Max. $\gamma$A & Max.
$\gamma\gamma$\\
&  & per Nucleon pair & (cm$^{-2}$s$^{-1}$ & Energy & Energy\\\hline
CERN SPS & Pb+Pb & 17 GeV & -- & 3.1 GeV & 0.8 GeV\\
RHIC & Au+Au & 200 GeV & $4\times10^{26}$ & 24 GeV & 6.0 GeV\\
RHIC & p+p & 500 GeV & $6\times10^{30}$ & 79 GeV & 50 GeV\\
LHC & Pb+Pb & 5.6 TeV & $10^{27}$ & 705 GeV & 178 GeV\\
LHC & p+p & 14 TeV & $10^{34}$ & 3.1 TeV & 1.4 TeV\\\hline
\end{tabular}
}
\end{center}
\end{table}}

\textrm{The equivalent (or virtual) photon flux per unit area (the relation
between $n(\omega)$ and $N(\omega, b)$ is $n(\omega)=\int N(\omega, b) d^{2}
b$) is \cite{Fe24,WW34,jackson}
\begin{equation}
N(\omega,b) = \frac{Z^{2}\alpha\omega^{2}}{\pi^{2}\gamma^{2} \hbar^{2}
\beta^{2} c^{2} } \bigg(K_{1}^{2}(x) + \frac{1}{\gamma^{2}} K_{0}%
^{2}(x)\bigg).\label{eq:ww1}%
\end{equation}
where $x=\omega b/\gamma\beta\hbar c$, $Z$ is the ion charge, $\alpha=1/137$,
$\beta c$ is the particle velocity and $K_{0}$ and $K_{1}$ are modified Bessel
functions. The first term ($K_{1}(x)^{2}$) gives the flux of photons
transversely polarized to the ion direction and the second is the flux for
longitudinally polarized photons. The transverse polarization dominates for
ultra-relativistic particles ($\gamma\gg1$). The photon flux is exponentially
suppressed when $\omega> \gamma\beta\hbar c/b$, justifying the estimates in
the beginning of this section. }

\textrm{These photons are almost real, with virtuality $- q^{2} < (\hbar
/R_{A})^{2}$. Except for the production of $e^{+}e^{-}$ pairs, the photons can
usually be treated as real photons. }

\textrm{The usable photon flux depends on the geometry. Most UPC reactions
lead to final states with a handful of particles. These final states will be
overwhelmed by any hadronic interactions between the fast moving ion and the
target. Thus, the useful photon flux is that for which the ions do not
overlap, \textit{i.e.} when the impact parameter $b=|\mathbf{b}_{1}
-\mathbf{b}_{2}|$ is greater than twice the nuclear radius ($2R_{A}$) (see
Figure \ref{diags}(c)). Usually, we can take $R_{A} = 1.2 A^{1/3}$ fm, where
$A$ is the atomic number. The $b>2R_{A}$ requirement treats the nuclei as hard
spheres; it is accurate for heavy nuclei, but less appropriate for lighter
ions. }

\textrm{The photons can interact with a target nucleus in a one-photon
process, (when $b_{1}<R_{A}$) or with its electromagnetic field in a
two-photon process when $b_{1}>R_{A}$ and $b_{2}>R_{A}$. In a photonuclear
(one-photon) interaction, the usable photon flux is obtained by integrating
Equation \ref{eq:ww1} over $b>b_{min}=2R_{A}$:
\begin{equation}
n(\omega) = {\frac{2Z^{2}\alpha}{\pi\beta^{2}}} \bigg[ \xi K_{0}(\xi)K_{1}%
(\xi) - \frac{\xi^{2}}{2}\big(K_{1}^{2}(\xi)-K_{0}^{2}(\xi
)\big)\bigg]\label{nofomega}%
\end{equation}
where $\xi=\omega b_{min}/\gamma\beta\hbar c = 2 \omega R_{A}/\gamma\beta\hbar
c$. }

\textrm{For two-photon exchange processes, the equivalent photon
numbers, in Equation \ref{eq:two-photon} must account for the
electric field orientation of the photon fluxes with respect to each
ion (see Figure \ref{diags}(b)), obeying the ion non-overlap
criteria $b_{1}, \ b_{2} >R_{A}$ \cite{BHTSK02}. The field
orientation is not included in Equation \ref{nofomega}. For
instance, owing to symmetry properties, $J^{\pi}=0^{+}$ (scalar)
particles originate from configurations such that $\mathbf{E}_{1}
\parallel \mathbf{E}_{2}$, whereas $0^{-}$ (pseudo-scalar) particles
originate from $\mathbf{E}_{1} \perp\mathbf{E}_{2}$
\cite{BF90,Vid93}. If one uses Equation \ref{nofomega} for
$n(\omega_{1})$ and $n(\omega_{2})$, the total photoproduction cross
section obtained from Equation \ref{eq:two-photon} is higher than in
a more detailed calculation, and the difference increases with
increasing particle masses \cite{BF90}. Even more detailed
calculations can be done by replacing the sharp-cutoff, $b_{1}, \
b_{2} >R_{A}$, criterion with integrals over $\mathbf{b}_{1}$ and
$\mathbf{b}_{2}$ which are weighted by the hadronic non-interaction
probability. Asymmetric collisions (especially $pA$ and $dA$) are
also of interest; the higher-$Z$ nucleus is likely to be the photon
emitter, so the photon direction is known. }

\textrm{Low energy processes, e.g. nuclear excitation are also
sensitive to the electromagnetic multipolarity involved. Equations
\ref{eq:ww1} and \ref{nofomega} are only appropriate for electric
dipole (E1) excitations. Equations for higher multipolarities are
described in Reference \cite{BB88}. }

\textrm{For protons, the hard sphere approximation is inadequate. Instead, the
proton size is included by the use of a form factor. With a dipole form
factor, the flux is \cite{DZ89}
\begin{equation}
n(\omega) = \frac{\alpha}{2\pi z} \big[1+ (1-z)^{2}\big] \bigg(\ln{\chi} -
\frac{11}{6} + \frac{3}{\chi} - \frac{3}{2\chi^{2}} +\frac{1}{3\chi^{3}%
}\bigg)\label{eq:ffluxp}%
\end{equation}
where
\begin{equation}
\chi= 1 + \frac{0.71 GeV^{2}}{Q_{min}^{2} c^{2}}%
\end{equation}
accounts for the proton structure and $z=W^{2}/s$, with $W$ the $\gamma p$
center of mass energy, and $s$ the squared ion-ion center of mass energy
per-nucleon. Here, $Q_{min}$ is the minimum momentum transfer possible in the
reaction. For proton-proton collisions, the form factor has an effect similar
to imposing a requirement $b_{min} = 0.7$ fm \cite{KN04}. }

\textrm{For protons and light nuclei, the weak electromagnetic interactions
introduce another complication. The momentum transfer due to elastic
scattering, $\Delta p = 2\eta\hbar/b$, with $\eta= Z_{1}Z_{2}\alpha\ll1$ is
small enough that the impact parameter is not a well defined observable
(because $\Delta p\Delta b \sim\hbar$, leading to $\Delta b > b$ for $\eta<
1$) \cite{multipleints}. This does not affect the total photon flux. However,
it might affect the component of the photon flux that is unaccompanied by
hadronic interactions. The uncertainty may also affect the probabilities for
multiple interactions, discussed in Section 5. }

\textrm{Equation \ref{eq:ffluxp} is valid when the proton remains intact. When
photon emission with proton excitation, such as to the $\Delta$ resonance, is
included, then the flux increases about 30\% \cite{OWZ94}. At very high photon
energy ($z\rightarrow1$), the magnetic form factor of the proton can also
become important \cite{K91}. }

\subsection{\textrm{Experimental Characterization}}

\textrm{Ultra-peripheral collisions look very different from the more
conventional hadronic interactions. The final state multiplicity is much
smaller, and, usually the events are fully reconstructed. Because the photon
$p_{T}$ are small ($p_{T}\approx\omega/\gamma c$), the final state $p_{T}$
will also be small. Photonuclear interactions that involve coherent scattering
from the target nucleus (such as vector meson production) also have a very
small $p_{T}$: $p_{T} < \hbar/ R_{A}$. This gives the events a distinctive
experimental signature, greatly simplifying detection \cite{STAR02}. }

\textrm{UPCs are studied at a variety of accelerators. The characteristics of
some relevant accelerators are given in Table~\ref{accelerators}. Each
accelerator can accelerate many different species; Table~\ref{accelerators}
gives only a few candidates. The CERN Super Proton Synchrotron (SPS) has
produced results on lead-to-bismuth transmutation and $e^{+}e^{-}$ pair
production in ion-ion collisions. }

\textrm{Although RHIC only began taking data in 2000, it has already released
UPC results on $\rho^{0}$ photoproduction and on $e^{+}e^{-}$ pair production.
RHIC has enough energy and luminosity to photoproduce a wide variety of
mesons, including the $J/\psi$. However, because it is a collider, detection
of very low $p_{T}$ particles is difficult, complicating the study of
$e^{+}e^{-}$ pairs and other atomic phenomena. }

\textrm{Although it is exclusively a $\overline p p$ collider, the Fermilab
Tevatron is an interesting place to study UPCs. Antihydrogen was produced
there using the process $\gamma\gamma\rightarrow e^{+}e^{-}$, with the
positron bound to an antiproton \cite{Blan98}. Photoproduction of the $J/\psi$
\cite{KN04} may have been observed by the CDF collaboration \cite{angela}. }

\textrm{The Large Hadron Collider (LHC), scheduled to begin operation in 2007,
will search for physics beyond the standard model. A UPC program at the LHC
can contribute to this search. Especially for $pp$ collisions, where
$\gamma\gamma$ and $\gamma p$ energies up to about 10\% of the beam energy are
accessible, UPCs may an attractive place to search for new physics. With ion
beams, the photon energies are lower, but $W^{\pm}$, $Z$, and heavy quark
physics may be studied. }

\section{\textrm{Low energy photonuclear interactions}}

\textrm{Relativistic Coulomb excitation (RCE) is now a popular tool to
investigate the intrinsic nuclear dynamics and structure of the colliding
nuclei. It is especially important in reactions involving radioactive nuclear
beams \cite{BCH93,Glas98,Au98,BP99,HJJ95,BHM02,BJ04}, and has been used for
many decades in low energy nuclear collisions to study nuclear structure
\cite{AW56}. However, nuclear-induced processes may also contribute to the
reactions being studied. }

\textrm{RCE may involve single or multiple photon-exchange between the
projectile and the target. In the first case, perturbation theory directly
relates the data to the matrix elements of electromagnetic transitions. These
matrix elements are clean probes of the nuclear structure, and RCE can be used
to study short-lived unstable nuclei that cannot be probed with real photons
or electron scattering \cite{Glas98,HJJ95,BJ04}. }

\textrm{Radiative capture processes ($b + c \rightarrow a +\gamma$) play a
major role in astrophysical sites, e.g., in a pre-supernova
\cite{Clayton,Rolfs}. Some reactions of interest for astrophysics, e.g.,
$^{7}Be\left(  p,\gamma\right)  ^{8}B$, can be studied via the inverse
photo-dissociation reaction $^{8}B\left(  \gamma,p\right)  ^{7}Be$
\cite{BBH86} using relativistic Coulomb collisions. The Coulomb breakup
reaction $a+A\longrightarrow b+c+A$ is useful to obtain the corresponding
$\gamma$ induced cross section $\gamma+a\longrightarrow b+c$. Using detailed
balance, this cross section can be related to the radiative capture cross
section $b+c\longrightarrow a+\gamma,$ of astrophysical interest \cite{BBH86}.
The radiative capture cross sections are often expressed in terms of the
astrophysical S-factor: $S(E)=\sigma(E)\ \exp(-2\pi Z_{b}Z_{c}e^{2}/\hbar
v_{bc})/E$, where $E\equiv E_{rel}=m_{bc}v_{bc}^{2}/2$ is the relative energy
between $b$ and $c$. In this equation $v_{bc}$ is the relative velocity and
$m_{bc}=m_{b}m_{c}/(m_{b}+m_{c})$ is the reduced mass of $b+c$. Because the
Coulomb penetration factor is explicitly factored out, the S-factor is a much
flatter function of $E$ than $\sigma(E)$ allowing a better extrapolation of
the measurements. }

\textrm{As an example, Figure \ref{join2}(a) shows the result of an experiment
performed at the GSI laboratory, in Darmstadt, Germany \cite{Sch04} for the
Coulomb dissociation of $^{8}$B. Data on the reaction $^{7}Be + p
\rightarrow\gamma+ ^{8}B$ is important for understanding the structure of our
sun. The decay of $^{8}B$ is responsible for the high energy neutrinos
observed by earth-bound detectors. The measured S-factor ($S_{17}$, 1=proton,
7=$^{7}$Be) is shown in Figure \ref{join2}(a) as solid circles. The solid
curve is a fit using a theoretical model for $S_{17}(E)$. Some of the data
shown in the figure are from direct capture experiments \cite{HammJ}. }

\textrm{\begin{figure}[tb]
\textrm{\centerline{\psfig{file=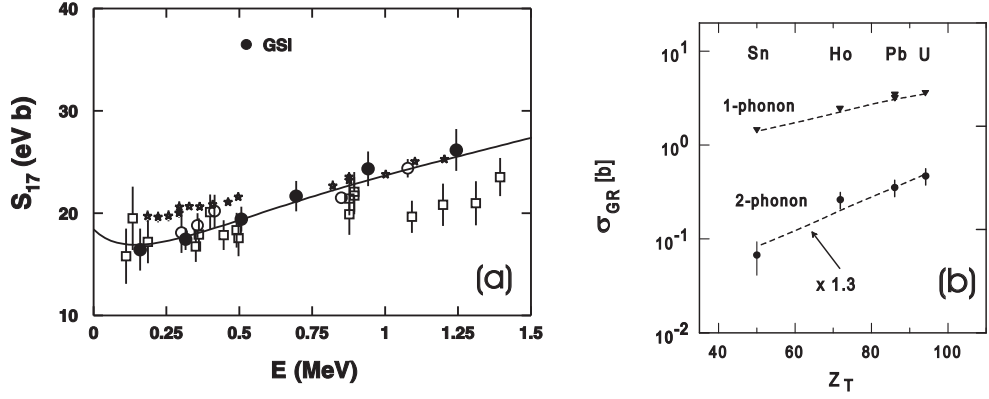,width=13cm}} \vspace*{8pt}
}\caption{{\small (a) S-factors ($S_{17}$) for the $^{7}$Be(p,$\gamma$)$^{8}$B
reaction. The GSI data was obtained using the Coulomb dissociation method
\cite{Sch04}. The other data are from direct capture measurements
\cite{HammJ}. (b) Cross sections for the excitation of the GDR (1-phonon) and
the DGDR (2-phonon) in $^{208}$Pb projectiles incident on different targets.
The dashed curves are theoretical calculations.}}%
\label{join2}%
\end{figure}}

\textrm{Other $b(c,\gamma)a$ radiative capture reactions are planned, or have
already been studied with the Coulomb dissociation method \cite{BR96}. These
processes may occur in the sun, supernovae, or during the Big Bang. Most of
these reactions cannot be directly studied because the Coulomb barrier leads
to very small values for the cross section, beyond the reach of present
experimental techniques. }

\textrm{
}

\textrm{Giant dipole resonances (GDR) occur in nuclei at energies of 10-20
MeV. Their gross properties are well described in terms of an out-of-phase
collective motion (oscillation) of protons against neutrons in a nucleus
\cite{GT48,SJ50}. If this oscillation is harmonic, with excitation energy
$\hbar\omega$, then higher excitation modes, with energies equal to
$N\hbar\omega$ also occur. These modes are interpreted as double, triple, ...,
giant dipole resonances. The double giant dipole resonances (DGDR) are thus
two giant dipole vibrations superimposed in one nucleus, with about twice the
energy of the GDR \cite{BB86,BB88,Au98,BP99}. In the harmonic model, the RCE
cross section for all multiphonon states can be calculated exactly
\cite{BB86}. }

\textrm{A series of experiments at the GSI laboratory obtained the energy
spectra, cross sections, and angular distribution of fragments following the
decay of the DGDR \cite{Ri93,Sc93,Au93,Bor96,Grun99,Bor99,Ili01}. The
experimental cross sections are about 30\% bigger than the theoretical ones.
This is shown in Figure \ref{join2}(b) where the dashed lines are the result
of theoretical calculations \cite{Au98,BP99,Bert96,HTV99,Tol99,Pau01,Psh01}.
These experiments are promising for the studies of the nuclear response in
very collective states. The giant resonances are still poorly understood, even
with the best current microscopic approaches. The study of the width of the
DGDRs will be helpful to improve this scenario \cite{BP99}. }

\textrm{In heavy ion colliders the mutual Coulomb excitation of the ions
(leading to their simultaneous fragmentation) is a useful tool for beam
monitoring \cite{seb1}. A recent measurement at RHIC \cite{seb2}, using the
Zero Degree Calorimeters to measure the neutron decay of the reaction
products, has proved the feasibility of the method. }

\textrm{DGDRs constitute only 10\% of the total fragmentation cross section
induced by Coulomb excitation in UPC. The dominant contribution is the
excitation of a single GDR, which then decays mostly by neutron emission. This
is also a major source of beam losses in relativistic heavy ion colliders
\cite{BB94}, and an important fragmentation mode of relativistic nuclei in
cosmic rays. }

\textrm{Another useful reaction is deuteron photodissociation in d+A
collisions - a photon from a heavy-ion photodissociates a deuteron
\cite{Hoff84}. The reaction has a large cross section, 1.38 b for d+Au at
RHIC, and 2.49 b for d+Pb at the LHC \cite{KV03}, and has been used as a
`standard candle' for luminosity monitoring \cite{SW04}. d+A collisions are
studied because they are technically simpler than p+A collisions. }

\section{\textrm{Photoproduction at Hadron Colliders}}

\textrm{The main interest in photoproduction at hadron colliders is derived
from the possibility it offers of a direct determination of the gluon
distribution in nucleons and nuclei. Examples of interactions in which the
gluon distribution can be probed are exclusive production of heavy vector
mesons, photoproduction of heavy quark-anti-quark pairs, and photoproduction
of jets. These gluon distributions are not directly accessible in deep
inelastic scattering, because the gluons carry neither electrical nor weak
charge. }

\textrm{Measuring the nuclear shadowing using heavy-ion beams is particularly
interesting. The nuclear gluon density can, as a first approximation, be
written as the nucleon gluon distribution, $g(x,Q^{2})$, multiplied by the
number of nucleons ($A$):
\begin{equation}
G^{A}(x,Q^{2}) = A g(x,Q^{2}) \; .
\end{equation}
Here, $x$ is the fraction of the projectile momentum carried by the gluon, and
$Q^{2}$ is the 4-momentum transfer squared. }

\textrm{Results from deep inelastic scattering of electrons on nuclear targets
have, however, showed deviations from such a simple scaling for the structure
function, $F_{2}(x_{2},Q^{2})$. Depending on $x$ and $Q^{2}$, suppression
(shadowing) of up to $\sim$30\% and enhancement (anti-shadowing) of up to
$\sim$10\% have been observed. The effects of shadowing on $G^{A}(x,Q^{2})$
are hard to determine directly. The current best estimates of the modification
to the gluon distribution in nuclei are obtained from the $Q^{2}$-evolution of
$F_{2}(x_{2},Q^{2})$\ \cite{Gousset,EKS98} and from studies of diffractive
interactions \cite{shadowdiffract}. Photoproduction at heavy-ion colliders may
provide a more direct measurement of $G^{A}(x,Q^{2})$. }

\textrm{The particle production in photon-hadron or photon-nucleus
interactions can be exclusive, when the protons or nuclei remain in their
ground state or are only internally excited, or inclusive, when at least one
of the nucleons or nuclei breaks up. Exclusive production will be discussed
first. When the momentum transfer is small compared with its inverse
nucleon/nuclear size, $Q \sim\hbar c/R$, the fields couple coherently to the
entire target. The kinematics of coherent, exclusive interactions is very
similar to that of two-photon interactions, which will be discussed in section
4. }

\subsection{\textrm{Exclusive Particle Production}}

\textrm{The dominant coherent interaction leading to the production of a
hadronic final state is the exclusive production of vector mesons,
\begin{equation}
A+A \rightarrow A+A+V .
\end{equation}
In these reactions a photon from the electromagnetic field of one of the
projectiles interacts coherently with the nuclear field of the other (target)
producing the vector meson. }

\textrm{Exclusive vector meson photoproduction on proton and nuclear targets
has been studied since the mid 1960's using photon beams \cite{THBauer}, and
since 1992 at the HERA electron-proton accelerator \cite{Crittenden}. The
first results from a heavy ion collider on exclusive $\rho^{0}$ production
($Au+Au \rightarrow Au+Au+\rho^{0})$ were recently published by the STAR
collaboration at RHIC \cite{STAR02}. }

\textrm{The total vector meson cross section in p+p or A+A interactions can be
calculated from Equation \ref{eq:photonuclear}. By differentiating and
changing variable from $\omega$ to $y$, the rapidity of the produced vector
meson, one obtains
\begin{equation}
\frac{d \sigma(A+A \rightarrow A+A+V)}{dy} = n(\omega) \sigma_{\gamma A
\rightarrow VA}(\omega)
\end{equation}
where the photon energy, $\omega$, is related to $y$ through $\omega= (M_{V}
c^{2}/2) \exp(y)$ and $M_{V}$ is the mass of the vector meson. If the photon
flux is known, the differential cross section, $d \sigma/dy$, is thus a direct
measure of the vector meson photoproduction cross section for a given photon
energy. }

\textrm{The bulk of the photon-hadron cross section can be explained by the
photon first fluctuating to a $q\bar{q}$ pair, which interacts with the target
through the strong nuclear force. Since the photon has quantum numbers $J^{PC}
= 1^{- -}$, it preferentially fluctuates to a vector meson. The lifetime of
the fluctuation is determined by the uncertainty principle. For a photon of
virtuality $Q$ fluctuating to a state of mass $M_{V}$ the lifetime is of the
order of
\begin{equation}
\Delta t \approx\frac{\hbar}{\sqrt{M_{V}^{2} c^{4} + Q^{2}c^{2}}} \approx
\frac{ \hbar}{M_{V}c^{2}} \ .
\end{equation}
The last approximation holds at hadron colliders because of the low virtuality
of the photons. The photon wave function is written as a Fock decomposition
\cite{SchSJ93}:
\begin{equation}
| \gamma> = C_{bare} | \gamma_{bare} > + C_{\rho} | \rho> + C_{\omega} |
\omega> + C_{\phi} | \phi> + ... + C_{q} | q \overline{q} >\ .
\end{equation}
Here $C_{bare} \approx1$ and $C_{V} \sim\sqrt{\alpha_{em}}$ ($V=\rho, \omega,
\phi, \cdots$). The coefficients $C_{V}$ are related to the photon-vector
meson coupling, $f_{V}$, through
\begin{equation}
C_{V} = \frac{\sqrt{4 \pi\alpha_{em}}}{f_{V}}.
\end{equation}
The numerical values of the couplings $f_{V}$ are usually determined from the
vector meson leptonic decay widths, $\Gamma(V \rightarrow e^{+} e^{-})$. }

\textrm{According to the Generalized Vector Meson Dominance Model (GVMD), the
scattering amplitude for the process $\gamma+ A \rightarrow B$ is the sum over
the corresponding vector meson scattering amplitudes,
\begin{equation}
A_{\gamma+A \rightarrow B}(s,t) = \sum_{V} C_{V} A_{V+A \rightarrow B}(s,t)
\;.
\end{equation}
For ``elastic'' scattering, $\gamma+A \rightarrow V+A$, the cross-terms, i.e.
$V^{\prime}+A \rightarrow V+A$, are usually small \cite{PautzShaw}, and are
often neglected. The cross section is then ($t$ is the momentum transfer from
the target nucleus squared and $d \sigma/dt = |A|^{2}$)
\begin{equation}
\frac{d \sigma(\gamma+A \rightarrow V+A)}{dt} = C_{V}^{2} \, \frac{d
\sigma(V+A \rightarrow V+A)}{dt} \; .
\end{equation}
}

\textrm{The momentum transfer of the elastic scattering is determined by a
hadronic form factor, $F(t)$,
\begin{equation}
\frac{d \sigma}{dt} = \left.  \frac{d \sigma}{dt} \right| _{t=0} | F(t) |^{2}
\; .
\end{equation}
For proton targets, the form factor is well represented by an exponential
function, $|F(t) |^{2}=\exp( - b |t|) $ with a slope $b \approx$%
~10~GeV$^{-2}~c^{2}$ for the light vector mesons ($\rho$, $\omega$) and $b
\approx$~4~GeV$^{-2}~c^{2}$ for the $J / \Psi$. The form factor for nuclear
targets is peaked at much smaller momentum transfers because of the larger
size of the target. }

\textrm{The form factor reflects the size and shape of the target. It can in
principle be calculated if the spatial distribution is known. The dynamical
information is contained in the forward scattering amplitude, $d \sigma/ dt
(t=0)$. The optical theorem relates this to the total vector meson cross
section, $\sigma_{tot}(VA)$:
\begin{equation}
\label{eq:optical}\left.  \frac{d \sigma}{dt} \right| _{t=0} = C_{V}^{2} \,
\frac{\sigma_{tot}^{2}(VA)}{16 \pi\hbar^{2} } (1 + \eta^{2}) \
\end{equation}
Here, $\eta$ is the ratio of the real to the imaginary part of the scattering
amplitude. }

\textrm{In Reference~\cite{KN99}, data on vector meson photoproduction with
proton targets were used to extract the total vector meson nucleon cross
section, $\sigma_{tot}(VN)$. This was then used to calculate the total vector
meson nucleus cross section, $\sigma_{tot}(VA)$, from the nuclear geometry.
This gave the vector meson production cross sections for heavy ion
interactions at RHIC and the LHC shown in Table~\ref{vm_sigma}. For heavier
vector mesons, like the $J/\psi$, gluon shadowing may reduce the cross-section
\cite{FSZ02a}. }

\textrm{\begin{table}[ptb]
\caption{Cross sections for exclusive vector meson production in Au+Au and
Pb+Pb interactions at RHIC and the LHC, respectively \cite{KN99}.}%
\label{vm_sigma}
\begin{center}
\textrm{%
\begin{tabular}
[c]{lcc}\hline\hline
Meson & Au+Au, RHIC & Pb+Pb, LHC\\
& $\sigma$ [mb] & $\sigma$ [mb]\\\hline
$\rho^{0}$ & 590 & 5200\\
$\omega$ & 59 & 490\\
$\phi$ & 39 & 460\\
$J/ \Psi$ & 0.29 & 32\\\hline\hline
\end{tabular}
}
\end{center}
\end{table}}

\textrm{In the Glauber Model \cite{Glauber59}, the elastic scattering
amplitude is given by the two-dimensional Fourier transform of the nuclear
profile function, $\Gamma(b)$:
\begin{equation}
\frac{d \sigma( \gamma+ A \rightarrow V+A ) }{dt} = \frac{\pi}{\hbar^{2}}
\left|  \int e^{i \mathbf{p}_{T} \cdot\mathbf{b}/\hbar} \, \Gamma(\mathbf{b})
\, d^{2}\mathbf{b} \right| ^{2}%
\end{equation}
$\Gamma(\mathbf{b})$ is a function of the distribution of matter inside the
nucleus, $\rho(\mathbf{b},z)$, and the vector meson-nucleon forward scattering
amplitude, $f_{VN}$ (which can be related to the total vector meson-nucleon
cross section through Equation~\ref{eq:optical}):
\begin{equation}
\Gamma( \mathbf{b} ) = 1 - \exp\left[  \frac{2i \pi\hbar c}{ \omega} \int\rho(
\mathbf{b} ,z^{\prime}) f_{VN}(0) dz^{\prime}\right] .
\end{equation}
This approach only works for high photon energies, when $c\gamma\beta\Delta t
>R_{A}$ so the interaction is longitudinally coherent over the entire nucleus.
At lower $\omega$, the loss of coherence reduces the cross section. The
Glauber model is discussed in References \cite{Grammer78} and \cite{Alberi81}.
}

\textrm{A Glauber model calculation of the coherent $\rho^{0}$ production
cross section in Au+Au collisions at RHIC gave a total cross section of 934~mb
\cite{FSZ02}. This is about 50\% higher than the result in \cite{KN99} (cf.
Table~\ref{vm_sigma}). The main reason for the difference is that in
\cite{KN99} the total vector meson-nucleus cross section was calculated
assuming that $\sigma_{tot}(\rho A) \approx\sigma_{inel}(\rho A)$. The
calculation in \cite{FSZ02} furthermore includes the contribution from
off-diagonal elements corresponding to $\rho^{\prime}+ Au \rightarrow\rho+ Au$
scattering, as well as a non-zero real part of the forward scattering
amplitudes ($\eta$ in Equation~\ref{eq:optical}). For a discussion of the
$\rho^{\prime}+ A \rightarrow\rho+ A$ contribution, see also \cite{PautzShaw}.
}

\textrm{The measured $\rho^{0}$ production cross section at RHIC is
$\sigma(Au+Au \rightarrow Au+Au+\rho^{0}) = 460\pm220\pm110$~mb at 130 A GeV
\cite{STAR02}. This can be compared with the Glauber Model calculations, which
give $\sigma=$~490~mb at this energy \cite{Frankfurt03}. The corresponding
number from the method used in \cite{KN99} is $\sigma=$~350~mb. }

\textrm{The rapidity distribution for coherent $\rho^{0}$ production measured
by the STAR collaboration in Au+Au interactions at 200 A GeV is shown in
\ref{star200_rho0}(a) \cite{Meissner03}. This is for $\rho^{0}$ production in
coincidence with mutual Coulomb breakup of the beam nuclei (cf. Section 5).
The rapidity distribution and cross section are in excellent agreement with
the distribution obtained from the Monte Carlo model based on the calculations
in \cite{BKN02}, corrected for the experimental acceptance. The reconstructed
invariant $\pi^{+} \pi^{-}$ mass is shown in Figure \ref{star200_rho0} (b).
The shape is well described by the sum of a relativistic Breit-Wigner function
and a S\"{o}ding interference term for direct $\pi^{+} \pi^{-}$ production
\cite{Soding66}. }

\textrm{\begin{figure}[tb]
\textrm{\centerline{\psfig{file=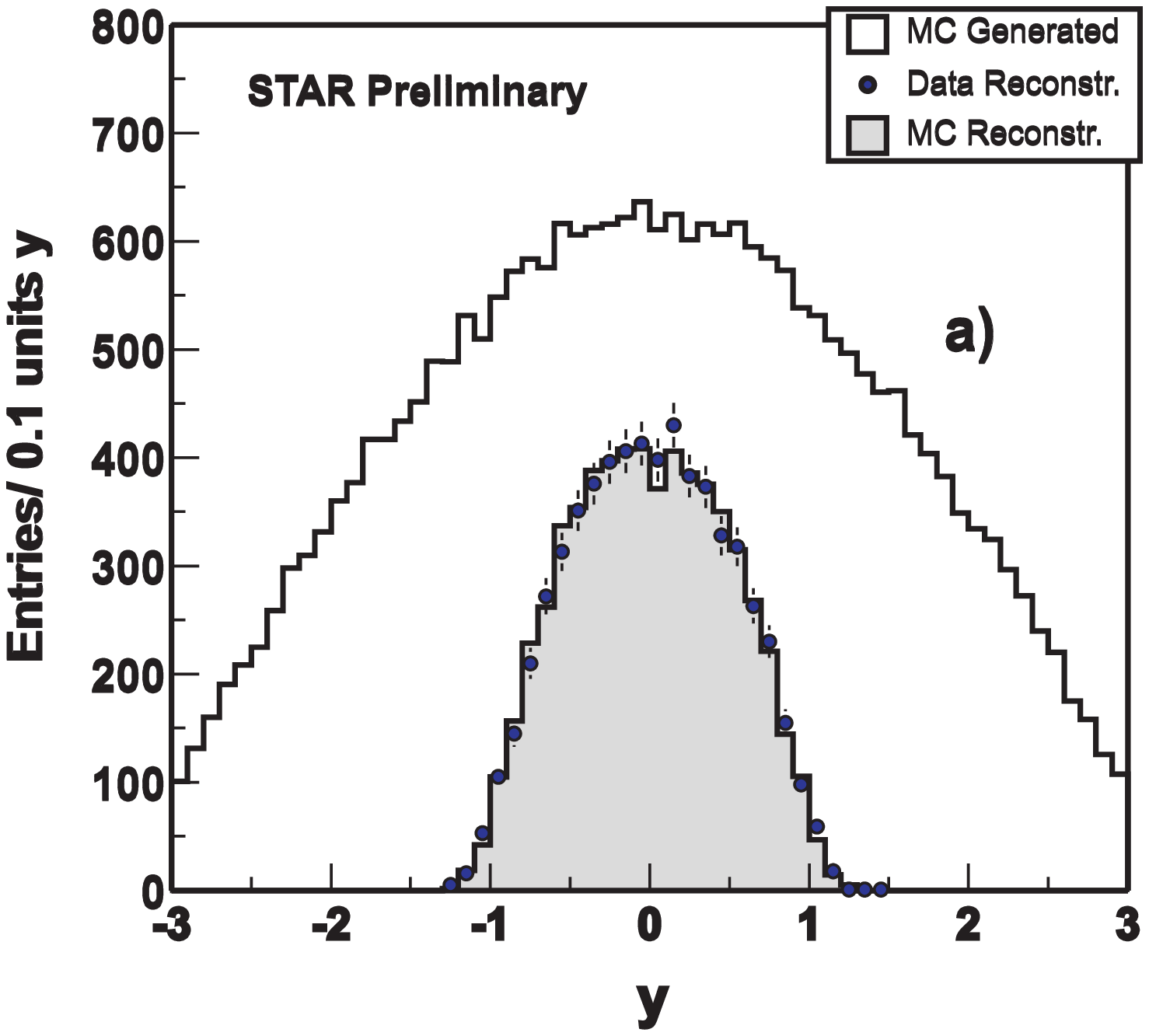,width=6 cm} \psfig{file=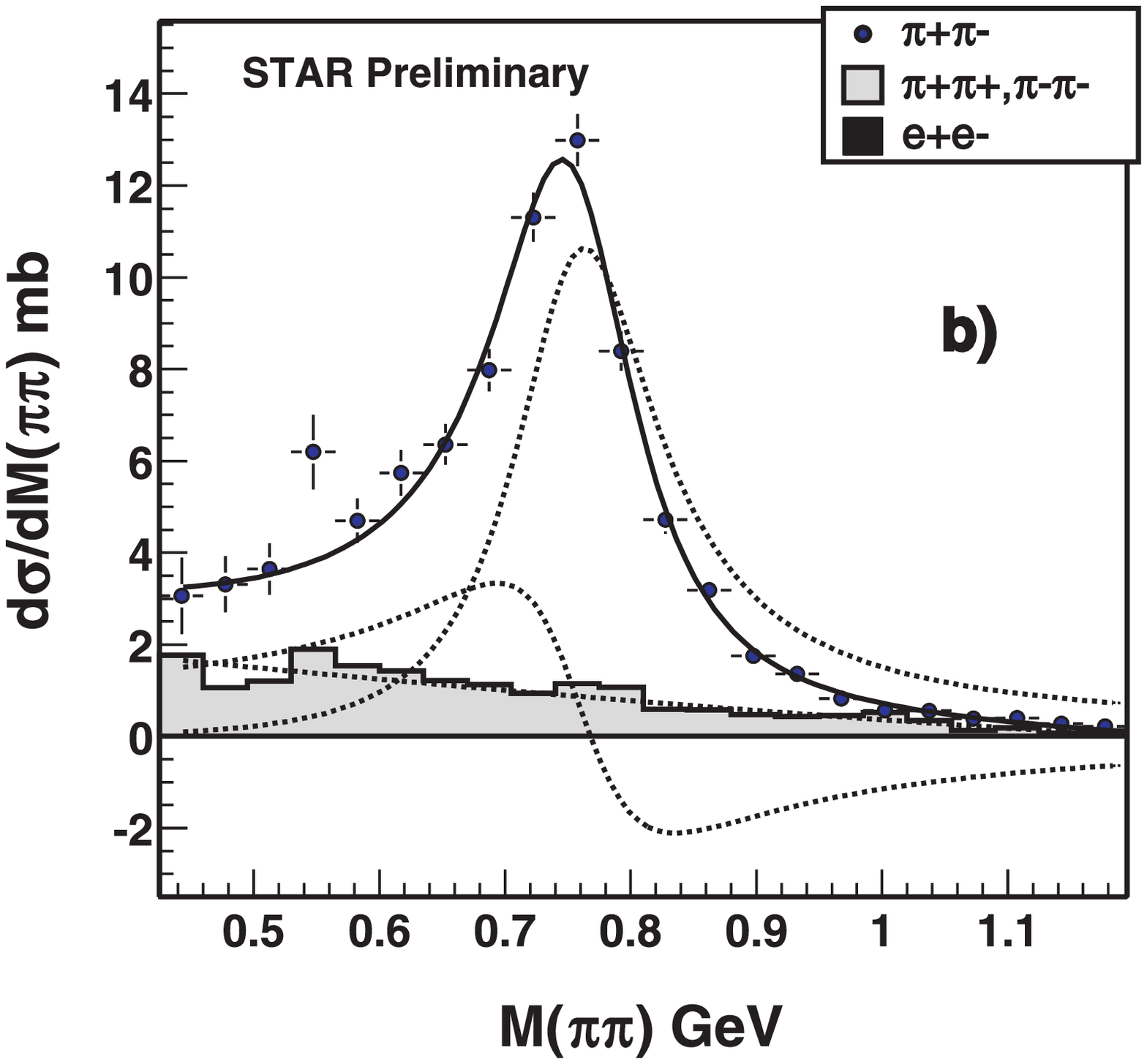,width=6 cm}}
\vspace*{8pt} }\caption{{\small Rapidity (a) and invariant mass (b)
distributions for coherent $\rho^{0}$ production in Au+Au interactions
accompanied by mutual Coulomb breakup at $\sqrt{s} = $ 200 A GeV, by the STAR
collaboration. The dashed curves in b) corresponds to a relativistic
Breit-Wigner function and a S\"{o}ding interference term; the solid curve is
the sum of the two. The dash-dotted curve describes the background from
incoherent interactions \cite{Meissner03}.}}%
\label{star200_rho0}%
\end{figure}}

\textrm{The STAR collaboration has very recently presented the first
preliminary data on $\rho^{0}$ production in deuteron-gold interactions (the
gold nucleus acts as photon emitter) \cite{Timoshenko05} and on coherent
production of $\pi^{+}\pi^{-}\pi^{+}\pi^{-}$ \cite{Ogawa05}; the latter may be
attributed to $\rho^{*}$ photoproduction. PHENIX has shown indications of
coherent $J/ \Psi$ and $e^{+} e^{-}$-pair production in Au+Au interactions at
RHIC \cite{SW04,Silvermyr04}. }

\textrm{The forward scattering amplitude for heavy vector mesons has been
calculated from two-gluon exchange in QCD. To leading-order \cite{Ryskin}
\begin{equation}
\left.  \frac{d \sigma(\gamma p \rightarrow Vp)}{dt} \right| _{t=0} =
\frac{\alpha_{s} ^{2} \hbar^{2} \Gamma_{ee}}{3 \alpha M_{V} ^{5} c^{6}} 16
\pi^{3} \left[  x g(x,M_{V}^{2}/4) \right] ^{2}.
\end{equation}
Here, $x$ is the fraction of the proton or nucleon momentum carried by the
gluons and the gluon distribution, $g(x,Q^{2})$, is evaluated at a momentum
transfer $Q^{2} = (M_{V}/2)^{2}$. This approach has been developed further by
including relativistic wave functions and off-diagonal parton distributions
\cite{Frankfurt99,Martin99}. The result is a total vector meson nucleon cross
section which grows rapidly with increasing photon-proton center-of-mass
energy, $W_{\gamma p}$. For $\Upsilon$ production, $\sigma\propto W_{\gamma
p}^{1.7}$ is expected. }

\textrm{The dependence of $d \sigma/dt$ on $[g(x)]^{2}$ makes exclusive vector
meson production a very sensitive probe of the proton and nuclear gluon
distributions. An $\Upsilon$-meson produced at mid-rapidity at the LHC would
come from gluons with $x \approx7 \cdot10^{-4}$ and $x \approx2 \cdot10^{-3}$
in p+p and Pb+Pb interactions, respectively. }

\textrm{Figure \ref{fig6} shows the predicted $d \sigma/dy$ for heavy vector
mesons in nucleus-nucleus and proton-proton collisions. The calculations are
based on parameterizations of the photon-proton cross sections derived from
measurements at HERA and from QCD based models \cite{KN04,KN99}. }

\textrm{\begin{figure}[tb]
\textrm{\centerline{\psfig{file=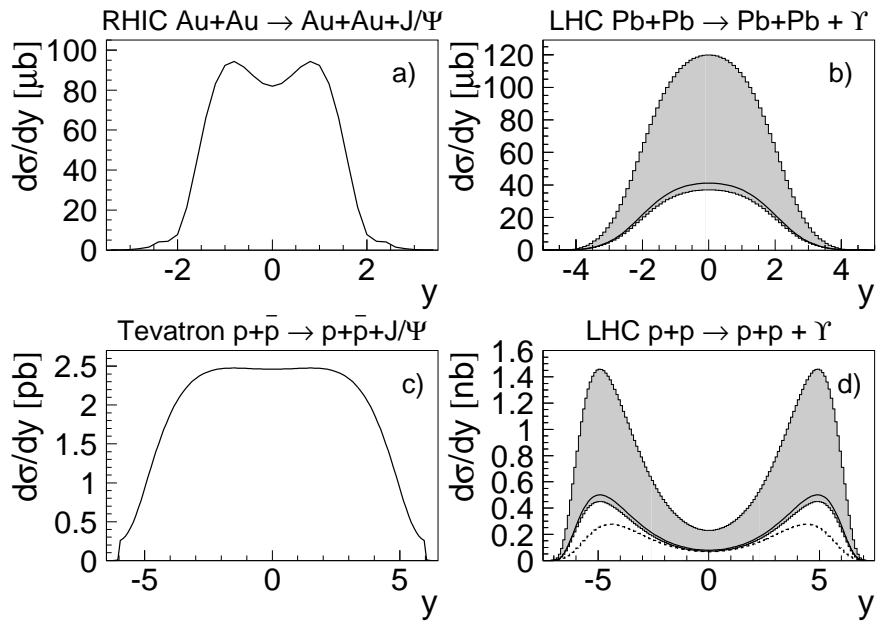,width=13cm}} \vspace*{8pt}%
}\caption{{\small Rapidity distributions for exclusive $J/\psi$ and $\Upsilon$
production in nucleus-nucleus and proton-proton collisions. Adapted from
\cite{KN04} and \cite{KN99}.}}%
\label{fig6}%
\end{figure}}

\textrm{The photoproduction cross sections rise rapidly with energy. With lead
at the LHC, the $\rho^{0}$ cross section is comparable to the hadronic cross
section. With calcium beams at full luminosity, the LHC will produce about
230,000 $\rho^{0}$, 15,000 $\phi$ and 800 $J/\psi$ per second. These rates are
comparable to dedicated $e^{+}e^{-}$ colliders, qualifying the LHC as a meson
factory! }

\subsection{\textrm{Interference in Exclusive Vector Meson Production}}

\textrm{When a single vector meson is produced through a coherent photonuclear
interaction in a nucleus-nucleus or proton-(anti-)proton collision, it is in
general not possible to determine which projectile acted as target and which
was the photon-emitter. The two possibilities are indistinguishable and under
certain conditions they will interfere quantum mechanically \cite{KN0003}.
Because of this interference, it is incorrect to add the cross sections for
the two possibilities. }

\textrm{The cross section is given by adding the corresponding amplitudes
$A_{1}$ and $A_{2}$
\begin{equation}
\label{interference}\frac{\hbar d \sigma}{dy dp_{T}} = \int_{b>2R} \left|
A_{1} \pm A_{2} \right| ^{2} d^{2} \mathbf{b} \; .
\end{equation}
The interference is maximal at mid-rapidity, where symmetry requires that
$A_{1}=A_{2}$. For ion-ion and proton-proton collisions, the interference is
destructive because of the negative parity of the vector meson; exchanging the
position of the two nuclei or the two protons is equivalent to a reflection of
the spatial coordinates, i.e. a parity transformation. For $p\overline p$
collisions, as at the Fermilab Tevatron, exchanging the proton and antiproton
involves a charge-parity (CP) transformation. Since CP is positive for vector
mesons, the interference is constructive at $p\overline p$ colliders
\cite{KN04}. }

\textrm{The amplitudes $A_{1}$ and $A_{2}$ depend on the photon flux
(\textit{e.g.} rapidity) and on the photonuclear cross sections. Their $p_{T}$
dependence comes from the convolution of the photon $p_{T}$ spectrum and the
$p_{T}$ from the photon-nucleus scattering. The former is given by the
equivalent photon spectrum \cite{Vid93,BF91}, and the latter comes from the
form factor of the target. }

\textrm{If the outgoing vector meson is treated as a plane wave (appropriate
for a distant observer), at mid-rapidity, $A_{1} = A_{2}$ and the square of
the sum of the amplitudes is
\begin{equation}
\label{costerm}\left|  A_{1} \pm A_{2} \right| ^{2} = 2 A_{0}^{2} \left(  1
\pm\cos( \frac{\mathbf{p} \cdot\mathbf{b}}{\hbar} ) \right)  \; .
\end{equation}
For very low momenta, $p_{T}\ll\hbar/ \langle b\rangle$, $\cos( \mathbf{p}
\cdot\mathbf{b} / \hbar) \approx1 - (\mathbf{p}\cdot\mathbf{b}/\hbar)^{2}/2$
and, as $p_{T} \rightarrow0$, the interference is complete; emission
disappears in ion-ion collisions, but doubles for $p\overline p$ colliders.
Interfence is significant for $p_{T} < 20$ MeV/c for the $\rho^{0}$ at RHIC
\cite{KN0003}, and $p_{T} < 250$ MeV/c for the $J/\psi$ at the Tevatron
\cite{KN04}. When $b\gg\hbar/p_{T}$, the cosine term oscillates rapidly as
\textbf{b} varies, and the interference disappears. In this regime, the cross
section reduces to the sum of cross sections for the two photon directions. }

\textrm{Away from mid-rapidity, $|A_{1}|\ne|A_{2}|$ because the photon
energies for the two possibilities are different: $\omega_{1,2} = (M_{V}
c^{2}/2)\exp(\pm y)$. Both the photon flux and photonuclear cross sections
will be different. The interference will thus be reduced. $A_{1}$ and $A_{2}$
could also have slightly different phases, adding a phase factor $\delta$ to
the cosine term in Equation \ref{costerm}. However, $\delta=0$ in the standard
Pomeron models \cite{KN0003}, and a significant phase difference seems
unlikely. }

\textrm{The interference in exclusive vector meson production is of particular
interest because the vector mesons have very short lifetimes compared to the
typical impact parameters. The median impact parameter for $\rho^{0}$
production in Au+Au collisions at RHIC, for example, is 46~fm, much larger
than the lifetime of the $\rho^{0}$ ($\tau$ = 1.3~fm/c). The vector meson
cannot, on the other hand, be produced more than $\sim$1~fm away from one of
the two nuclei because of the short range of the nuclear force. Observing the
expected interference pattern would thus prove that the wave function of the
vector meson is preserved long after it has decayed. This is an example of the
Einstein-Podolsky-Rosen paradox \cite{KN0003}. }

\textrm{Preliminary data from the STAR collaboration on $\rho^{0}$ production
in Au+Au at $\sqrt{s_{nn}}$ = 200~GeV seem to confirm the presence of
interference \cite{SRK0304}. The measured $t$ spectrum is shown in
Figure~\ref{star_int}. The data are for interactions where both gold nuclei
Coulomb dissociate. The coincident Coulomb dissociation selects events with
smaller impact parameters compared to exclusive production (cf. Section 5);
with Coulomb dissociation, the median impact parameter is only 18 fm
\cite{BKN02}. }

\textrm{\begin{figure}[t]
\textrm{\centerline{\psfig{file=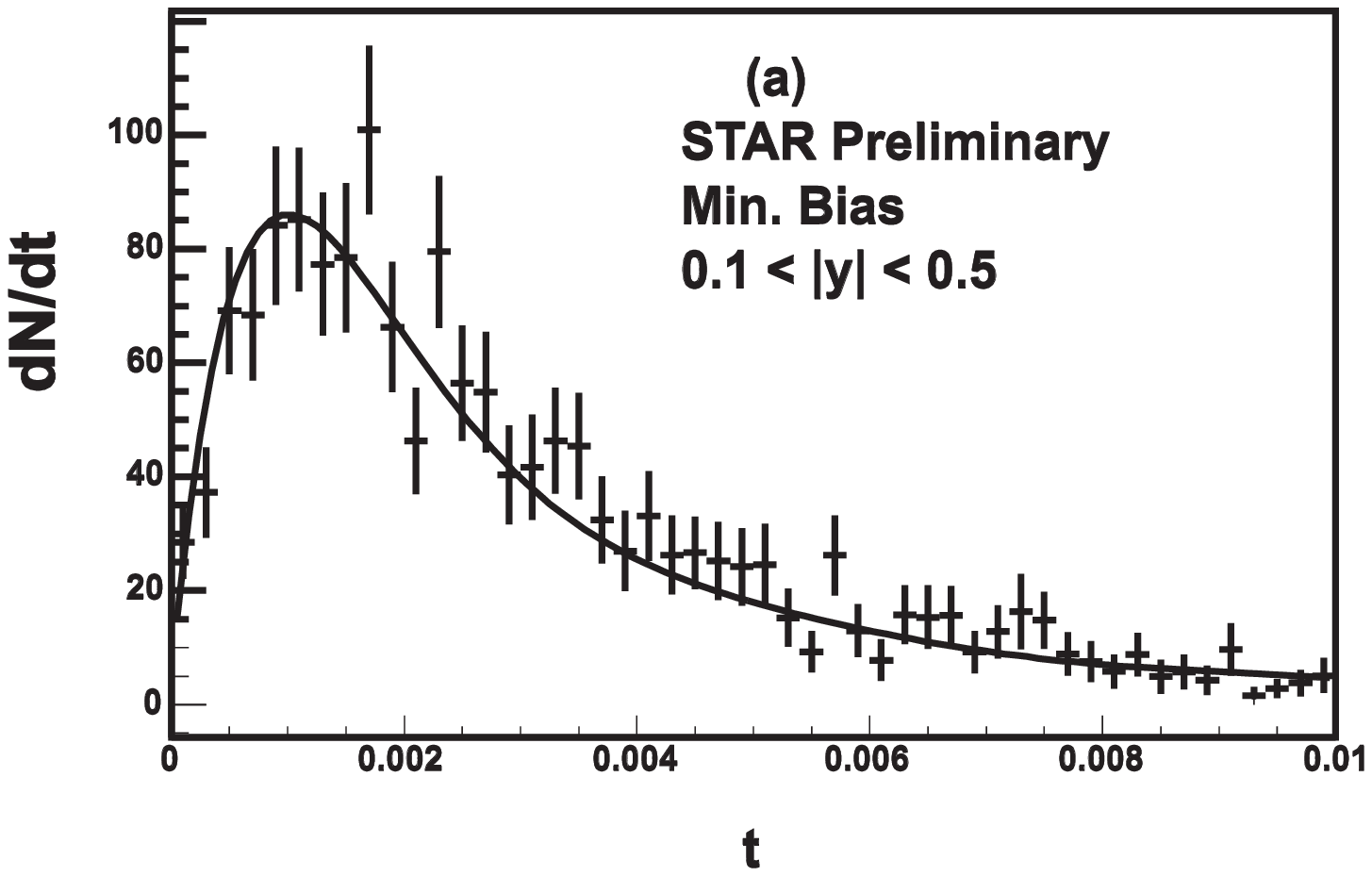,width=7cm} \psfig{file=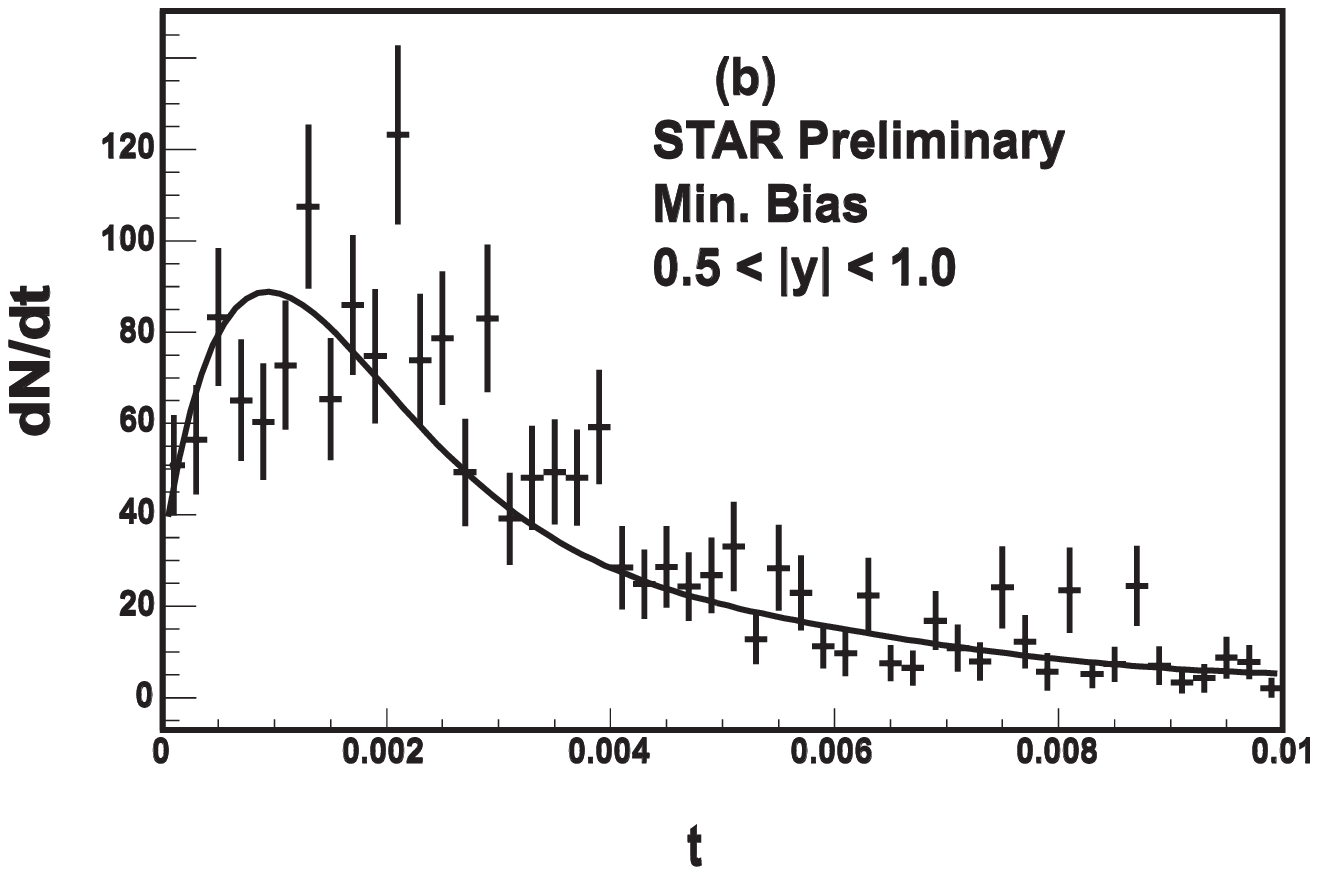,width=6.7cm}}
\vspace*{8pt} }\caption{{\small Efficiency corrected $t_{\perp}$ spectrum for
$\rho^{0}$ production in Au+Au collisions at RHIC with Coulomb breakup. The
points are the data, and the solid curve is a fit to
Equation~\ref{eq:interference}. Interference causes the dip at low $t \leq$
0.001~$(GeV/c)^{2}$ \cite{SRK0304}.}}%
\label{star_int}%
\end{figure}}

\textrm{The data are fit to a function,
\begin{equation}
\label{eq:interference}\frac{dN}{dt} = a \, \exp( - b |t| ) \, \left[  1 + c
(R(t)-1) \right] ,
\end{equation}
with three parameters. These correspond to a normalization constant ($a$), the
width of the nuclear form factor ($b \approx R_{A}^{2}/\hbar^{2}$), and a
parameter to quantify the magnitude of the interference ($c$). The function
$R(t)$ is the ratio of Monte Carlo $d\sigma/dt$ calculated with and without
interference. This functional form separates the interference from the nuclear
form factor. No interference would correspond to $c=0$, while complete
interference according to the calculations above would correspond to $c=1$. A
fit to the data finds $c=1.01 \pm0.08$ ($0.1 \leq y \leq0.5$) and $c=0.78
\pm0.13$ ($0.5 \leq y \leq1.0$) for the two ranges in rapidity. }

\subsection{\textrm{Inclusive Photoproduction}}

\textrm{The high photon flux at hadron colliders and the large total
photon-hadron cross sections lead to high rates for other photonuclear
interactions. In Au+Au collisions at RHIC, the total photonuclear cross
section for photon-nucleon center-of-mass energies above 4 GeV is about 2
barns, or nearly 1/3 of the total hadronic Au+Au cross section. The majority
of these interactions are resolved interactions, i.e. they are preceded by a
fluctuation of the photon to a $q \overline{q}$ state. They therefore resemble
inelastic hadron-nucleon/nucleus collisions. Because the photon energies are
much lower than the beam energies, the kinematics is similar to that of fixed
target interactions. }

\textrm{Despite the large cross section, particle production in resolved
photon-nucleon/nucleus interactions at hadron colliders has so far attracted
relatively little interest. See, however, Reference \cite{Pshenichnov99}.
Understanding the kinematics of these interactions is nevertheless essential,
because they are a significant background to other ultra-peripheral processes,
particularly at the trigger level \cite{KN98}. }

\textrm{Considerably more interest has been devoted to direct photon
interactions, in particular the production of heavy $q \overline{q}$-pairs
\cite{GoBe02,GVHSS95,KNV01,KNV02}. Recently, the cross section for
photoproduction of heavy quark pairs in pp collisions has been calculated
\cite{Goncalves}. In these interactions, a photon interacts with a parton in
the target and the partonic cross section can be calculated from QCD. }

\textrm{The leading order contribution to the photoproduction of a $q
\overline{q}$-pair corresponds to photon-gluon fusion, as is illustrated in
the Feynman-diagram in Figure~\ref{diags} (d). The cross section for the
partonic sub-process is \cite{JonesWyld,FriStreng78}
\begin{equation}
\sigma_{\gamma g \rightarrow q \overline{q}} \, (W_{\gamma g}) = \frac{\pi
e_{q}^{2} \alpha_{em} \alpha_{s}(Q^{2}) \hbar^{2} c^{2}}{W_{\gamma g}^{2}}
\left[  ( 3 - \beta^{4} ) \ln(\frac{1 + \beta}{1 - \beta}) -2 \beta(2 -
\beta^{2}) \right]  \; .\label{lo_qqbar}%
\end{equation}
Here, $m_{q}$ and $e_{q}$ are the quark mass and electric charge,
respectively, $\beta= (1 - 4 m_{q}^{2} c^{4}/W_{\gamma g}^{2})$ and $W_{\gamma
g}$ is the photon-gluon center-of-mass energy. If the gluon carries a fraction
$x$ of the nucleon momentum, then $W_{\gamma g}^{2} = 2 \omega x \sqrt{s}$.
The strong coupling constant, $\alpha_{s}(Q^{2})$, is evaluated to one loop at
scale $Q^{2} = m_{q}^{2} c^{2} + p_{T}^{2}$, where $p_{T}$ is the quark
transverse momentum. }

\textrm{The total photoproduction cross section $\sigma( A[\gamma]A
\rightarrow A q \overline{q} X)$ is obtained by convoluting the partonic cross
section with the equivalent photon flux, $n(\omega)$, and the nuclear/nucleon
gluon distribution, $G^{A}(x_{2},Q^{2})$, i.e.
\begin{equation}
\sigma( A[\gamma]A \rightarrow A q \overline{q} X) = \int\int\frac{n(\omega
)}{\omega} \, G^{A}(x,Q^{2}) \, \sigma_{\gamma g}(W_{\gamma g}) \,
\Theta(W_{\gamma g} - 2 m_{q} c^{2}) \, d \omega dx \, .\label{aa_qqbar}%
\end{equation}
This equation is the equivalent of Equation~\ref{eq:two-photon} for two-photon
interactions with the photon flux from one nucleus replaced by the gluon
distribution, $G^{A}(x,Q^{2})$. The final state $q\overline q$ rapidity
depends on the photon energy and the gluon $x$. The rapidity distributions of
bottom and top quarks produced in Pb+Pb and O+O collisions at the LHC are
shown in Figure~\ref{qqbar_rapidity}. The kinematics are discussed in more
detail elsewhere \cite{KNV02,SmNe}. }

\textrm{\begin{figure}[t]
\textrm{\centerline{\psfig{file=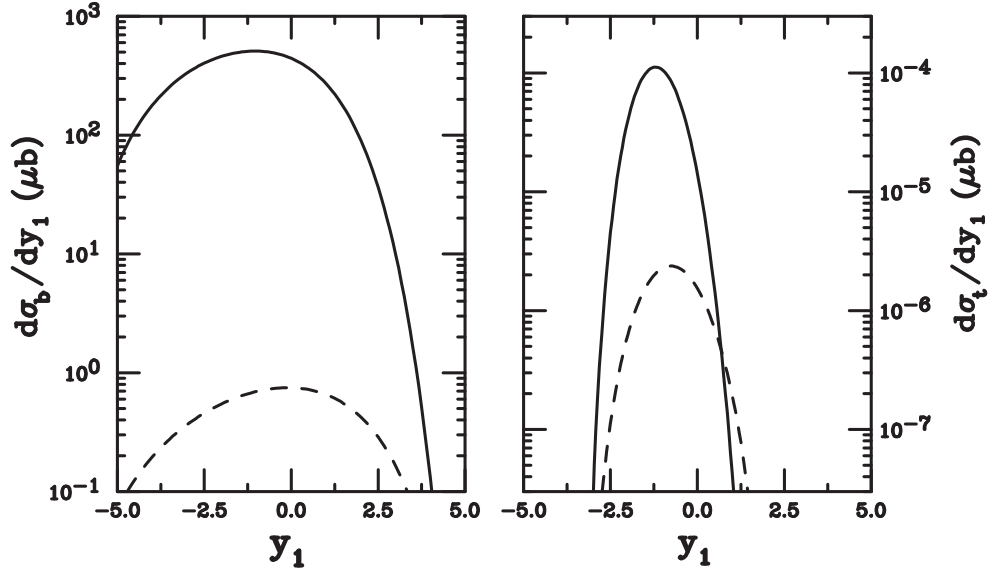,width=13cm}} \vspace*{8pt}
}\caption{{\small Rapidity distributions of bottom (left) and top (right)
quarks produced in photonuclear collisions at the LHC. The solid and dashed
curves are for Pb+Pb and O+O interactions, respectively. Here, the photon is
emitted by the nucleus with positive rapidity; the complete cross section is
the sum of this curve plus it's mirror image. Shadowing is included using the
parameterizations of Reference~\cite{EKS98}. Adapted from References
\cite{KNV01} and \cite{KNV02}.}}%
\label{qqbar_rapidity}%
\end{figure}}

\textrm{The $q \overline{q}$ production cross section is peaked near
threshold, $W_{\gamma g} \approx4 m_{q}^{2}$. Mid-rapidity production of $c
\overline{c}$- and $b \overline{b}$-pairs therefore mainly probes x-values of
$x \sim1 \cdot10^{-3}$ ($c \overline{c}$) and $x \sim3 \cdot10^{-3}$ ($b
\overline{b}$) in heavy-ion collisions at the LHC. The corresponding numbers
at RHIC are $x \sim2 \cdot10^{-2}$ and $x \sim1 \cdot10^{-1}$. (For a $q
\overline{q}$-pair with invariant mass $W_{\gamma g}$ and pair-rapidity $y$,
$x = (W_{\gamma g}/\sqrt{s}) e^{y}$.) }

\textrm{The total cross sections for $c \overline{c}$ and $b \overline{b}$
production in various systems at RHIC and the LHC are listed in
Table~\ref{gamglu_sigma}. The calculations without shadowing are compared with
two calculations that include nuclear modifications. As expected, shadowing
has the largest effect on the production of lighter quarks ($c \overline{c}$)
using heavy nuclei. The cross section for producing a $c \overline{c}$-pair in
Pb+Pb interactions at the LHC is of the order of 1~b, nearly 1/6 of the total
hadronic cross section. }

\textrm{\begin{table}[tb]
\caption{Cross sections for $q \overline{q}$ photoproduction through direct
photon-gluon fusion in heavy ion interactions. The numbers in column 4 and 5
include nuclear gluon shadowing from References \cite{EKS98} and \cite{FGS},
respectively.}%
\label{gamglu_sigma}
\begin{center}
\textrm{%
\begin{tabular}
[c]{lcccc}\hline\hline
\multicolumn{5}{c}{$q \overline{q}$ cross sections in heavy ion collisions}%
\\\hline
Colliding system & flavor & $\sigma$ [mb] & $\sigma$ [mb] & $\sigma$ [mb]\\
&  & No shadowing & EKS98 & FGS\\\hline
RHIC Au+Au & $c \overline{c}$ & 15.8 & 17.4 & 17.6\\
RHIC Au+Au & $b \overline{b}$ & 2.9 $\cdot10^{-3}$ & 3.0 $\cdot10^{-3}$ & 3.0
$\cdot10^{-3}$\\
RHIC Si+Si & $c \overline{c}$ & 0.196 & 0.203 & 0.20\\
RHIC Si+Si & $b \overline{b}$ & 1.07 $\cdot10^{-4}$ & 1.13 $\cdot10^{-4}$ &
1.14 $\cdot10^{-4}$\\
LHC Pb+Pb & $c \overline{c}$ & 1250 & 1050 & 850\\
LHC Pb+Pb & $b \overline{b}$ & 4.9 & 4.7 & 4.4\\
LHC Ar+Ar & $c \overline{c}$ & 16.3 & 14.3 & 12.3\\
LHC Ar+Ar & $b \overline{b}$ & 0.073 & 0.070 & 0.066\\\hline
\end{tabular}
}
\end{center}
\end{table}}

\textrm{Quark pairs can also be produced in anomalous interactions, where a
parton from the resolved photon interacts with a parton in the target, or in
two-photon interactions. The cross sections from anomalous interactions are
small compared with the direct production cross sections \cite{KNV02}. The
anomalous cross sections are 1-20\% of the direct cross sections, depending on
quark flavor and collision energy. The two-photon contribution is usually less
than 1\% of the anomalous cross section. }

\textrm{In addition to probing the nuclear gluon distribution, the
photon-gluon fusion reactions are of interest as a means to determine the
electric charge of the top quark. The top quark measurements are all
consistent with a standard model top with electric charge $q_{t} = +2/3$.
However, since the correlation between the decay products $W^{+}
\leftrightarrow b$ or $W^{-} \leftrightarrow\overline{b}$ are never measured,
$q_{t}$ is unconstrained \cite{BBO01}. The data can still accommodate an
exotic particle $t^{\prime}$ with $q_{t} = -4/3$, which decays via $t^{\prime
}\rightarrow W^{-} + b$. Since the cross section for $\gamma+ g \rightarrow q
+ \overline{q}$ is proportional to $q_{t}^{2}$, $q_{t} =$ -4/3 would quadruple
the cross section. }

\textrm{There have so far been no experimental measurements of heavy quark
pair photoproduction in heavy ion or pp collisions. Some of the experimental
techniques that could be used to separate this signal from hadronic
backgrounds are discussed in Section 6. }

\subsection{\textrm{Dijets, Compton scattering, Vector Boson Production and
other processes}}

\textrm{When the final state quarks in the process shown in Figure~\ref{diags}
(d) have high $p_{T}$, the final state is two roughly back-to-back jets
\cite{dijets}. The cross section is calculated in a manner similar to that of
Equation \ref{lo_qqbar}, except that light quarks are included. Jets may also
be produced to leading order through the process $\gamma+q \rightarrow g + q$.
The cross section for dijet production is sensitive to the gluon distribution
in the target; because of the simplicity of the reaction, there are fewer
systematic uncertainties than in other processes, like vector meson
production. However, it may be more difficult to isolate dijet photoproduction
from backgrounds, such as hadronic production of dijets and diffractive
hadronic production. }

\textrm{Since the gluon-contributing nucleus does not stay intact, the
experimental signature for this process is two jets, accompanied by a single
rapidity gap between the jets and the photon-emitting nucleus. The two jets
may have very different rapidities and it may be difficult to reconstruct the
entire event. Calculations have considered the case where a single jet is
detected, with $|\eta|<1$ \cite{dijets}. Without shadowing, the rate to
photoproduce jets with energies above 21 GeV in lead-lead collisions at the
LHC is 0.015 Hz. In a $10^{6}$ s run, jets up to 80 GeV should be detectable.
}

\textrm{A closely related process is the production of a photon + jet final
state; this is essentially Compton scattering. The rates for this process are
about two orders of magnitude below that for dijet production \cite{dijets}. }

\textrm{The strong Coulomb fields may also dissociate hadrons into jets. For
example, a proton may fragment into three quarks, leading to reactions such as
$\gamma p\rightarrow3\ jets$; this would be a distinctive signature in $pA$
collisions \cite{FS03}. One photon can also dissociate another, leading to
reactions such as $\gamma\gamma\rightarrow2\ jets$ \cite{FS03,FELIX}. }

\textrm{$W^{\pm}$ and $Z^{0}$ can be photoproduced in ultra-peripheral
collisions. A $Z^{0}$ can be produced when a photon fluctuates to a
$q\overline q$ pair, scatters from a target nucleus and emerges as a $Z^{0}$.
In the high-energy limit, the cross section for $\gamma p\rightarrow Z^{0} p$
is about 0.01 pb \cite{pumplin}. Unfortunately, even with a coherent photon
beam, the cross section seems too low to be observable. }

\textrm{$W^{\pm}$ pairs can be produced directly from a photon fluctuation,
$\gamma\rightarrow W^{+}W^{-}$. One of the $W$s can interact with the target
nucleus, leading to a hadronic jet plus a real $W$. Unfortunately, this
process has not been studied in detail. }

\section{\textrm{Two-Photon Processes}}

\subsection{\textrm{Production of free- and bound-pairs}}

\textrm{Between 1933 and 1937, Furry, Carlson, Landau, Lifshitz, Bhabha,
Racah, Nishina, Tomonaga, and several others performed calculations of
$e^{+}e^{-}$ production in relativistic collisions of fast particles (cosmic
rays) \cite{Fu33,LL34,Bh35,Nis35,Rac37}. The purpose was to test the newly
born Dirac theory for the positron. Starting with the Dirac equation for the
electron and its antiparticle they found \cite{Rac37},
\begin{equation}
\sigma={\frac{28}{27\pi}} \sigma_{0} \left[  L ^{3} - 2.198 L ^{2}+3.821 L
-1.632\right]  ,\label{LL}%
\end{equation}
where $\sigma_{0}=({Z_{1}Z_{2}\alpha^{2} \hbar/ m_{e} c})^{2}$, $L=\ln
(\gamma_{1}\gamma_{2})$, and $\gamma_{i}$ is the Lorentz factor of ion $i$ in
the laboratory system. The first term of this equation can be simply obtained
from Equation 1 and the cross sections for $\gamma\gamma$ pair-production. The
production cross sections for heavy lepton pairs ($\mu^{+}\mu^{-}$, or
$\tau^{+}\tau^{-}$), can be obtained similarly. The electromagnetic production
of $\mu^{+}\mu^{-}$ pairs using hadron beams was first observed in 63 GeV $pp$
collisions at CERNs Intersecting Storage Rings \cite{Ant80}. }

\textrm{For meson pairs like $\pi\pi$, neglecting internal substructure, as is
done for Equation \ref{LL} may be appropriate near threshold. However, at
higher pair masses, the quark substructure of the mesons becomes important,
and the cross section for $\gamma\gamma\rightarrow\pi^{0}\pi^{0}$ becomes
comparable to that for $\pi^{+}\pi^{-}$ \cite{MPW94}. In fact, studies of
$\gamma\gamma$ production of mesons pairs are interesting probes of meson
structure. Baryon-antibaryon pairs are also of interest, because the reaction
probes the baryon internal structure. }

\textrm{Because the cross sections depend on the inverse of the square of the
particle mass, production of heavier pairs ($\mu^{+}\mu^{-}$, $\tau^{+}%
\tau^{-}$) is much smaller than for $e^{+}e^{-}$. Their Compton wavelength,
$\lambda_{i}= \hbar/m_{i} c$ is smaller than the nuclear radius $R$. This
requires the replacement $L\rightarrow\mathcal{L}=\ln(\gamma_{1}\gamma
_{2}\delta/m_{i}cR)$ in Equation \ref{LL}, where $\delta=0.681...$ is a number
related to Eulers constant. }

\textrm{Bound particles, such as positronium or $q\overline q$ mesons are also
produced in two-photon interactions. The cross section is given by Equation 2.
The cross section for $\gamma_{1}\gamma_{2}\rightarrow X$ depends on the
particle's decay width to two photons, $\Gamma_{\gamma\gamma}$ \cite{Low60}.
Since decay and $\gamma\gamma$ production use the same matrix elements, only
the phase-space factors and polarization summations are distinct. One finds
\cite{Low60}
\begin{equation}
\sigma(\omega_{1}, \ \omega_{2} ) = 8\pi^{2} (2J+1){\frac{\Gamma_{\gamma
\gamma}}{Mc^{2}}} \ \delta(4\omega_{1} \omega_{2} - M^{2}c^{4})\label{Low}%
\end{equation}
where $J$, $M$, and $\Gamma_{M\rightarrow\gamma\gamma}$ are the spin, mass and
two-photon decay width of the meson. The delta-function imposes energy
conservation. }

\textrm{Using Equation \ref{Low}, the production of mesons with mass $M$ in HI
colliders is \cite{BB88}:
\begin{equation}
\sigma= {\frac{128}{3}} \ \left( Z_{1}Z_{2}\alpha\right) ^{2} {\frac
{\hbar\Gamma_{\gamma\gamma}}{M^{3}c^{5}}} \left[  \mathcal{L}^{3}
+\cdots\right]  \ .\label{ln}%
\end{equation}
This equation is obtained by using Equation \ref{eq:two-photon} and the high
energy limit ($\gamma\gg1$) of the equivalent photon number $n(\omega)$ (for
more details, see reference \cite{BB89}). }

\textrm{A more detailed account of the space geometry of the two-photon
collision is necessary \cite{BF90}, especially for heavier mesons, and will be
discussed in Section 4.4. Since spin 1 particles cannot couple to two real
photons \cite{Ya50}, only spin 0 and spin 2 particles should be produced. }

\textrm{The treatment of bound states in quantum field theory (QFT) is a very
complex subject (for reviews, see \cite{BYG85,Sap90}). In the case of
positronium production by two photons (para-positronium) and by three photons
(ortho-positronium), standard QFT techniques allow a simple and accurate way
to calculate the cross sections from first principles \cite{KKSS99,BeNa02}.
The para-positronium production cross sections are quite large, 19.4 mb and
116 mb, for RHIC (Au+Au) and LHC (Pb+Pb), respectively \cite{BeNa02}. However,
Coulomb corrections reduce these values by as much as 43\% for RHIC and 27\%
for LHC \cite{KKSS99}. The cross section for the production of
ortho-positronium, which requires three-photon exchange, are also large: 11.2
mb and 35 mb, for RHIC and LHC, respectively \cite{KKSS99}. Even the
ortho-positronium cross sections correspond to production rates of 4 and 35
per second respectively. If the positronium could be extracted from the the
interaction points, they could be used to test interesting properties of QFT
for bound states. Relativistic positronium has an unusual large transparency
in thin layers (see \cite{postr} and references therein). }

\textrm{The same diagrams for the calculation of the positronium apply for
production of bound $q\bar q$ pairs (mesons) in UPC \cite{BeNa02}. However,
proper accounting for the color degrees of freedom is needed \cite{App75}. }

\subsection{\textrm{Production of free $e^{+}e^{-}$ pairs}}

\textrm{Due to experimental difficulties, Equation \ref{LL} (and its newer
counterparts) has never been fully tested. With the construction of RHIC and
the LHC, interest in this process has grown. For heavy ions, the $e^{+}e^{-}$
production probabilities are close to one and lowest-order perturbative
calculations of the cross sections violate unitarity (\textit{i.e.}
$d^{2}\sigma/d^{2}b >1$) \cite{BB88}. }

\textrm{This observation lead to more detailed calculations
\cite{Bau90,RB91,RhW91,BGS92,Vid93,HTB95} involving high-order processes, such
as the exchange of multiple photons (Coulomb distortion) and the production of
multiple pairs, as shown in Figures \ref{diags}(e,f). These processes are
important for collisions at small impact parameters. Diverse theoretical
methods have been considered. Perturbative calculations are simple to write
down, but they involve rather complicated integrals, especially for low energy
electrons, due to Coulomb distortion and relativistic effects on the continuum
electronic wavefunction \cite{BB88}. A general sum of the contribution of
diagrams like those in Figure \ref{diags}(e,f) and unitarity corrections
(involving the production of virtual $e^{+}e^{-}$ pairs) was obtained in
Reference \cite{LMS02}. To account for Coulomb distortions, one needs to add
to eq. \ref{LL} a term of the form (see eq. 7.3.10 of Reference \cite{BB88}),
\begin{equation}
\sigma_{C}=-{\frac{28}{9 \pi}}\left[ f(Z_{1}\alpha)+f(Z_{2}\alpha)\right]
\sigma_{0}L^{2} \ ,\label{CC}%
\end{equation}
where
\begin{equation}
f(x)=x^{2}\sum_{n=1}^{\infty}{\frac{1}{n(n^{2}+x^{2})}} \
\end{equation}
is the Bethe-Maximon correction. Eq. \ref{CC} was derived in Ref. \cite{Iv99},
and later confirmed by independent calculation in Ref. \cite{Lee00}. }

\textrm{For Pb+Pb collisions at LHC the Coulomb distortion correction reduces
the pair-production cross section by 14\%. Other unitarity corrections further
reduce the cross sections by 3\% \cite{LMS02}. }

\textrm{The calculation of the production of multiple pairs, as shown in
Figure \ref{diags}(e,f), is directly connected with the unitarity problem. It
is possible to interpret $d^{2}\sigma/d^{2}b$ as the mean number of pairs
produced at a given impact parameter. For Ca-Ca collisions at the LHC
($Z\alpha= 0.15$), $\sigma_{2-pairs} = 0.11$ b \cite{LMS02}, or about 27 000
$e^{+}e^{-}e^{+}e^{-}$ events per second. In the literature, one finds
different methods to calculate the cross section for $n>2$ pairs. The result
of Ref. \cite{Guc00} is a simple fit to numerical calculations. Ref.
\cite{Bau90} is based on the probability $P(b)$ of the pair production taken
from Ref. \cite{BB88}. Ref. \cite{LMS02} claims that this expression is wrong.
They derived expressions for this probability using two different methods.
Reference \cite{LMS02} obtains $\sigma_{n} \propto L^{n}$, whereas References
\cite{Bau90} and \cite{Guc00} obtain $\sigma_{n} \propto L^{3n}$ respectively.
The calculations differ in the method used to include Coulomb and unitarity
corrections. The production of multiple pairs has been studied with a variety
of different theoretical approaches.
\cite{Iv99,LMS02,Ast02,ERSG99,HTB99,BHTSK02,Ei00,HBT04,BGMP01}. }

\textrm{The calculation of multiple photon exchange can be considerably
simplified in the ultra-relativistic limit. In this limit, the electromagnetic
field of the ions is squeezed in the plane perpendicular to its trajectory
(i.e. it can be approximated by a delta-function along this plane). In the
appropriate gauge, the Coulomb potential is two-dimensional and the
time-dependent Dirac equation may be solved exactly \cite{Ba91,Baltz95,Ba97}.
This should be equivalent to an all-orders perturbative calculation. }

\textrm{This approach yields good results as long as $\omega b/\gamma\hbar c
\leq0.1$ \cite{Ber01}. Above this value the delta-function approximation
breaks down. Since the most important impact parameters for this process are
of the order of \ $b \simeq\hbar/m_{e} c$ \cite{BB88}, the calculation can be
separated into two regions: (a) $b \simeq\hbar/m_{e} c$ where the
approximation is valid, and (b) large impact parameters, for which
perturbative calculations are accurate \cite{Ba91,Baltz95,Ba97}. This method
describes well the differential cross sections for $e^{+}e^{-} $ pair
production up to energies of the order of $0.1\gamma m_{e} c^{2}$, above which
the delta-function approximation breaks down for the same reason as above
\cite{Ber01}. The initial calculations using this technique found results that
matched the lowest order perturbation theory \textit{without} Coulomb
corrections \cite{BM98,SW98}. This was inconsistent with both theoretical
expectations and with data \cite{SW98}. However, regularization of the
integrals was critical; with regularization, the Dirac approach reproduced the
lowest order result, \textit{with} Coulomb corrections \cite{Lee00,Baltz04}.
This technique allows for the calculation of cross section for free
$e^{+}e^{-}$ pair production to all orders in $Z\alpha$ \cite{Baltz04}. }

\textrm{Electron-positron pair production has been studied at RHIC in
combination with mutual Coulomb excitation \cite{STAR04}. As will be discussed
in the next section, the mutual Coulomb excitations were required to trigger
on these events, and also had the effect of selecting events with $b<30$ fm,
where non-perturbative effects were strongest. The cross section, pair mass,
rapidity and pair $p_{T}$ distributions were all in accord with the
predictions of lowest order perturbation theory \cite{HBT04}. The pair $p_{T}$
distribution deviated from the Weiszaecker-Williams virtual photon prediction,
showing that the photon mass was important in that kinematic regime. }

\textrm{The STAR study suffered from low statistics. Earlier experiments on
$e^{+}e^{-}$ production in sulfur-platinum (and sulfur-nuclei) collisions at
the CERN SPS had higher statistics, but lower beam energies; they also found
good agreement with lowest order calculations \cite{CERNee,CERNee2}. }

\textrm{It remains disappointing that these 70-year-old QED calculations are
still not fully tested. Although many aspects of QED have been tested to high
precision, studies involving strong fields are much less advanced.
Pair-production with relativistic heavy ions (with $Z\alpha\sim1$) is one
important example. }

\subsection{\textrm{Pair production with capture and antihydrogen}}

\textrm{An important phenomenon occurs when the electron is captured in an
atomic orbit of the projectile, or of the target \cite{BB88}. At RHIC and the
LHC, this is an important source of beam particle loss \cite{BB89}. The
produced beam of single-electron ions carries considerable power
\cite{beampipe}; at the LHC, at full luminosity, the produced $^{+81}Pb$ beam
carries sufficient power to quench the superconducting accelerator magnets;
this limits the LHC luminosity with heavy-ions \cite{LHCbeams}. }

\textrm{One striking application of this process was the recent production of
antihydrogen atoms using relativistic antiproton beams \cite{Bau96}. The
positron is produced and captured in an orbit of the antiproton. The
expression
\begin{equation}
\sigma=3.3\pi{\frac{Z_{1}^{2}Z_{2}^{6}\alpha^{8}\hbar^{2}}{m_{e}^{2}}}%
\ {\frac{1 }{\exp\left(  2\pi Z_{2}\alpha\right)  -1}} \left(  L -2.051\right)
\label{eq:bound}%
\end{equation}
for pair production with electron capture in the nucleus with charge $Z_{2}$
is obtained in first order perturbation theory \cite{BB88}. Although Equation
\ref{eq:bound} works reasonably well for explaining antihydrogen production,
it is only valid for small $Z_{i}$ ($Z_{i}\leq15$) \cite{BD01,Bau96,BB98}. For
large $Z_{i}$, as with the experiments at RHIC and LHC, non-perturbative
calculations may be necessary \cite{AB87,Bec87,RB91,RB90,Ba97,He00}. Equation
\ref{eq:bound} includes higher order effects related to the electron capture,
but is not a complete all-orders result. The additional higher order
corrections are apparently small, and Equation \ref{eq:bound} should be usable
for most purposes. The fraction with the exponential term is due to the
distortion of the positron wavefunction. It accounts for the reduction of the
magnitude of the positron (continuum) wavefunction near the nucleus where the
electron is localized (bound). }

\textrm{Equation \ref{eq:bound} shows that the cross section depends on energy
as $\sigma=A\ln\gamma_{1} \gamma_{2}+B$, where the coefficients $A$ and $B$
depend on the nuclear charges. This scaling was confirmed in numerical
calculations of Reference \cite{Ba97} and was used in the analysis of the
experiment in Reference \cite{Kr98}, shown in Figure \ref{join3}(a). The
comparison between theory and the data of Reference \cite{Kr98}, is not
completely valid since atomic screening was not included. When screening is
present the cross sections are smaller by at least a factor 2-4 (see Equation
7.4.3 of Reference \cite{BB88}). }

\textrm{\begin{figure}[t]
\textrm{\centerline{\psfig{file=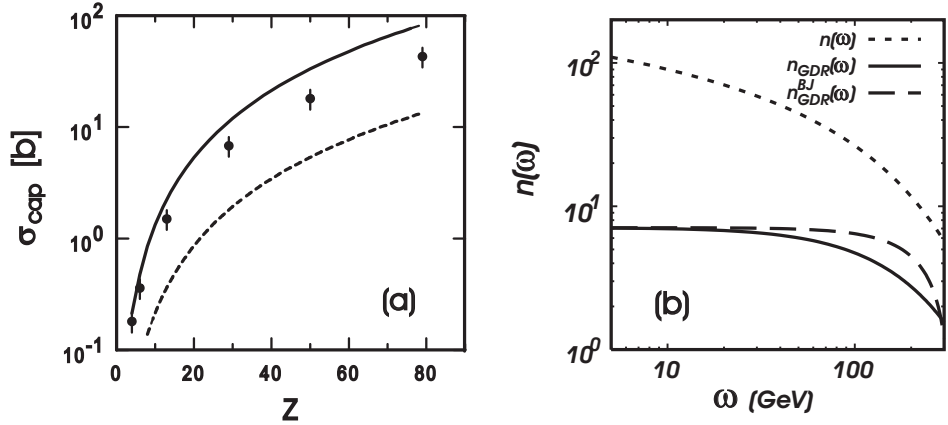,width=13cm}} \vspace*{8pt}
}\caption{ {\small (a) Pair production with capture for $Pb^{82+}$ (33 TeV)
beams on several targets. The solid and dashed curves are theoretical
calculations \cite{Kr98}. See text for more details. (b) The tagged total
photon flux (accompanied by a single giant dipole resonance excitation) for a
complete calculation (solid line) and a simplified 'box' integration (dashed
line), compared to the untagged flux (dotted line), for gold at RHIC. Although
tagging reduces the low-energy flux by an order of magnitude, at high
energies, the difference is much smaller. From Reference \cite{multipleints}%
.}}%
\label{join3}%
\end{figure}}

\textrm{Similar reactions occur with hadron pairs. Processes like pion pair
production with capture (or, similarly, photoexcitation of a $\Delta$
resonance, which decays by $\pi^{+}$ emission) increase the nuclear charge by
1 and can turn lead into bismuth, plus a $\pi^{+}$. This change cannot occur
electromagnetically, since $e^{+}$ will not bind to lead. These transmutations
have been studied at the SPS/CERN for $^{208}$Pb at 160 GeV/nucleon
\cite{Sch02}. The data can be described quantitatively with electromagnetic
excitation calculations \cite{Pshenichnov99,Pshe}. For high-Z nuclei, the
dominant contribution to nuclear-charge pickup is due to electromagnetic
production of $\pi^{-}$ by virtual photons. The electromagnetic contribution
is completely negligible for a similar experiment at an energy of 10.6 GeV per
nucleon \cite{Ger95}. }

\subsection{\textrm{Two-photon production of mesons}}

\textrm{For the production of composite particles, one can use Equation
\ref{ln} as a first guess. However, even the lightest mesons ($\pi^{0}$)
require photons of relatively large energy ($\geq70$ MeV). Mesons are produced
primarily in collisions with relatively small impact parameters (compared to
$2R_{A}$) where hard photons are more abundant. Substantial changes are
required to Equation \ref{ln} to account for the collision geometry
\cite{CJ90,BF90}. }

\textrm{One can rewrite Equation \ref{eq:two-photon} more conveniently as
\begin{equation}
\sigma_{X}=\int ds {\frac{d\mathcal{L}(W_{\gamma\gamma})}{dW_{\gamma\gamma}}}
\ \sigma_{\gamma\gamma\rightarrow X}(W_{\gamma\gamma}) \ ,\label{lumin}%
\end{equation}
where $W_{\gamma\gamma}=4\omega_{1}\omega_{2}$ is the square of the
center-of-mass energy of the two photons, $\sigma_{\gamma\gamma\rightarrow X}
(W_{\gamma\gamma})$ is the two-photon production of particle X, and $d
\mathcal{L}/dW_{\gamma\gamma}$ is the ``photon-photon luminosity".
$d\mathcal{L}/dW_{\gamma\gamma}$ can be multiplied by the ion beam
luminosities, yielding an ``effective" two-photon luminosity $d\mathcal{L}%
_{eff}/dW_{\gamma\gamma}$ which can be directly compared to other two-photon
luminosities, such as at $e^{+}e^{-}$ or pp colliders \cite{Khoze02}. With
heavy ion beams, the LHC two-photon luminosities are much higher than are
available elsewhere, either with proton beams at the LHC or at the $e^{+}%
e^{-}$ LEP-II collider for energies up to $\sqrt{W_{\gamma\gamma}}\approx500$
GeV \cite{BHTSK02}. }

\textrm{Table \ref{ggmesons} \cite{BeNa02} shows the cross sections for the
production of $C=even$ mesons for the RHIC and LHC colliders. Other
calculations were done in References \cite{KN98,FELIX,GVS93}. The cross
section corresponds to a $\gamma\gamma\rightarrow\pi^{0}$ rate of 30
events/second with lead beams at the LHC. For heavier mesons, like $\eta_{c}$,
the rate is still large, of the order of 1 per minute. }

\textrm{\begin{table}[tb]
\caption{Cross sections for two-photon production of ($C=even$) mesons at RHIC
(Au+Au) and at LHC (Pb+Pb) \cite{BeNa02}.}%
\label{ggmesons}%
\textrm{{%
\begin{tabular}
[c]{lcccccl }\hline\hline \toprule meson & mass [MeV] &
$\sigma^{RHIC}$ [mb] & $\sigma^{LHC}$ [mb] \cr \colrule $\pi_{0}$ &
134 & 4.9 & 28\\\hline $\eta$ & 547 & 1.0 & 16 &  &  & \\\hline
$\eta^{\prime}$ & 958 & 0.75 & 21 &  &  & \\\hline $f_{2} (1270)$ &
1275 & 0.54 & 22 &  &  & \\\hline $a_{2} (1320)$ & 1318 & 0.19 & 8.2
&  &  & \\\hline $\eta_{c}$ & 2981 & 3.3$\times10^{-3}$ & 0.61 &  &
& \\\hline $\chi_{0c}$ & 3415 & 0.63$\times10^{-3}$ & 0.16 &  &  &
\\\hline
$\chi_{2c}$ & 3556 & 0.59$\times10^{-3}$ & 0.15 &  &  & \\
\botrule &  &  &  &  &  &
\end{tabular}
} }\end{table}}

\textrm{For mesons of comparable mass, the two-photon cross sections in
Table~\ref{ggmesons} are about two orders of magnitude lower than the cross
sections for photonuclear vector meson production (Table~\ref{vm_sigma}). This
difference stems from the different coupling strengths of the strong and
electromagnetic interactions, $\alpha_{s} \sim1$ and $\alpha_{em}
\approx1/137$. }

\textrm{Two-photon meson spectroscopy is thus greatly complicated by the large
background from photonuclear interactions. For example, with lead beams at the
LHC, the rate of $J/\psi$ photoproduction followed by $J/\psi\rightarrow
\gamma\eta_{c}$ is about 2.5 per minute, higher than the $\gamma
\gamma\rightarrow\eta_{c}$ rate. }

\textrm{Although it may be possible to separate the different event classes
with cuts on meson $p_{T}$, rapidity, and final state particles, the vector
meson background seems daunting to most efforts \cite{KN99}. }

\subsection{\textrm{Searches for New Physics}}

\textrm{The LHC will reach high enough energies that two-photon interactions
will be an attractive place to search for some types of new physics. Many
early calculations focused on the search for the Higgs \cite{Pap89,
Grab89,Dree89}. Other examples include supersymmetric particle pairs, magnetic
monopoles and possible extra spatial dimensions \cite{BHTSK02}. The LHC will
also be able to probe vector boson couplings through reactions like
$\gamma\gamma\rightarrow W^{+}W^{-}$. }

\textrm{The two-photon production rate for the Higgs is small enough that, for
most models, it is likely to be discovered in hadronic interactions. However,
for standard model Higgs masses under $\approx200$ GeV, with medium ion beams,
the $\gamma\gamma$ channel should produce a handful of events per year
\cite{BHTSK02}. In some supersymmetric scenarios, the production of the Higgs
in UPCs could be significantly enhanced \cite{GVS93}. The $\gamma\gamma$
production channel could also be studied in $pp$ collisions; the greatly
increased luminosity and running time will more than compensate for the
smaller production cross section. However, in $pp$ collisions, there is a
considerable background due to diffractive interactions. It may be possible to
separate $\gamma\gamma$ from diffractive interactions by studying the $p_{T}$
of the scattered protons. This could be done by placing small detectors, known
as Roman pots inside the beampipe, to detect protons scattered at very small
angles \cite{Pio01}. }

\textrm{Supersymmetric particle pairs are similar story. If present, they are
likely to be discovered in hadronic interactions. However, if supersymmetry is
correct, a large number of new particles are likely to be present, and the
$\gamma\gamma$ production of sparticle pairs is likely to provide significant
new information; two photon production is sensitive to significant regions of
phase space \cite{GVS93}. Two-photon interactions are sensitive to sparticles
that do not participate in the strong interaction. Photonuclear interactions
may also be useful for studying supersymmetry. }

\textrm{Real or virtual magnetic monopoles can be produced by two-photon
fusion. The CDF collaboration searched for the process $\gamma\gamma
\rightarrow\gamma\gamma$ wat the Fermilab Tevatron, and set mass limits on
magnetic monopoles with various charges \cite{CDF98}. The LHC will be able to
do far better. }

\textrm{The presence of extra dimensions could be detected via the two-photon
production of gravitons. The cross section to produce a graviton,
$\gamma\gamma\rightarrow G$ increases in the presence of compact dimensions
\cite{ANP00}. There are unresolved theoretical issues regarding the cross
section, but one calculation finds that, for 2 extra compact dimensions, at
the LHC, the cross section is of order 1 nb for lead and 10 pb for calcium
\cite{ANP00}. Rates are not given for proton beams, but, for calcium, and
likely protons, a few events would be produced each month. The experimental
signature of graviton production has not been worked out in detail. }

\textrm{Most of these `new physics' channels involve relatively high $p_{T}$
particles, and so should be within the purview of the planned trigger setup
for the ATLAS and CMS detectors. This may not be true for supersymmetric final
states; two charged sleptons that don't interact hadronically will challenge
any trigger. }

\section{\textrm{Multiple Interactions Between a Single Ion Pair}}

\textrm{Because heavy-ions have such large charges, a single ion pair can
undergo multiple electromagnetic reactions in a single interaction. Even
though the reactions may be independent, the geometry introduces correlations
between the photon energies and polarizations. Multiple interactions are also
a key experimental tool, allowing for many cross checks under different
triggering conditions. One example of such a reaction is the photoproduction
of a $\rho^{0}$ meson, accompanied by the mutual Coulomb excitation of both
nuclei:
\begin{equation}
Au + Au \rightarrow Au^{*} + Au^{*} + \rho^{0}.\label{rhostarstar}%
\end{equation}
This reaction was studied by the STAR collaboration \cite{STAR02}. This
process occurs predominantly via 3-photon exchange, as is shown in Figure
\ref{diags}(g). }

\textrm{The STAR collaboration has also observed four photon reactions, such
as the production of an $e^{+}e^{-}$ pair accompanied by mutual Coulomb
excitation \cite{STAR04}, as is shown in Figure \ref{diags}(h). }

\textrm{In a multi-photon process, each photon emission may be treated
independently, if the energy lost by the nucleus is not significant. As long
as the photon emission does not excite the emitter, the reactions may be
treated as completely independent. The cross section is calculated in impact
parameter space
\begin{equation}
\sigma= \int d^{2}b P(b)\label{eq:multipleints}%
\end{equation}
where $P(b)$ is the probability for the reaction to occur at impact parameter
$b$. This is
\begin{equation}
P(b) = \int{\frac{d\omega}{\omega}} N(\omega,b) \sigma_{\gamma A}%
(\omega).\label{eq:pofb}%
\end{equation}
When the cross section for a reaction is very large (as with $e^{+}e^{-}$
production or GDR excitation) the naive $P(b)$ calculated in Equation
\ref{eq:pofb} may exceed 1. $P(b)$ should then not be interpreted as a
probability but rather as the mean number of produced particles at that impact
parameter. The actual number of produced particles follows a Poisson
distribution with this mean. The generalization to 2 (or more) photon
exchanges is obvious. Calculations using this approach accurately predicted
the cross section and kinematic distributions for Reaction \ref{rhostarstar}
\cite{STAR02}. }

\textrm{This factorization only holds if several conditions are satisfied.
Photon emission must not excite the emitting nucleus and the photons must be
emitted independently. As long as the fractional energy loss of the nuclei are
small, this is valid \cite{Gup55}. Finally, the excitation must not change
nuclear form factor significantly on the time scale of the reaction (the
`frozen nucleus' approximation). As long as these strictures are satisfied,
the ordering of the subprocesses is unimportant. These conditions hold for
heavy-ion collisions. For proton beams, with $\eta=Z_{1}Z_{2}e^{2}\alpha
/\beta\ll1$, the factorization is on weaker ground, because of the poorly
defined impact parameter \cite{multipleints}. }

\textrm{It can be convenient to treat one reaction as a trigger (or selector)
for a range of impact parameters. Picking events with mutual Coulomb
excitation, for example, selects events with small impact parameters. The
reason can be seen in Equation \ref{eq:multipleints}. In the low-energy limit
($\omega\ll\gamma\hbar c/R_{A}$), for a fixed $\omega$, $P(b)\approx
N(\omega,b) \approx1/b^{2}$. For a two-photon reaction, $P_{1}(b)P_{2}%
(b)\approx1/b^{4}$. The mean impact parameter $b_{n}$ for an n-photon
interaction is \cite{multipleints}
\begin{equation}
\overline b_{n} = \frac{\int d^{2}b b P_{1}(b)...P_{n}(b)} {\int d^{2}b
P_{1}(b)...P_{n}(b)}.
\end{equation}
Here, $b_{min} = 2R_{A}$ is the minimum impact parameter, and $b_{max} =
\gamma\hbar/R_{A}$. For $n=1$ the result $\overline b = (b_{max}-b_{min}%
)/\ln{(b_{max}/b_{min})}$ is not so useful. However, for 2 or more photon
exchanges
\begin{equation}
\overline b_{n>1} = 2R_{A} \frac{2n-2}{2n-3};
\end{equation}
$b_{max}$ drops out, leaving a simple result. For $n=2$, this reduces to
$\overline b_{2}=2R_{A}$; for larger $n$, $\overline b$ is even smaller. At
heavy-ion colliders, mutual Coulomb excitation is an effective trigger for
selecting low-impact parameter events. Detailed calculations of the median
impact parameter in 1 and 3 photon interactions find a similar scaling
\cite{BKN02}. Reducing $\overline b$ is very helpful in studying interference
in vector meson production, by increasing the $p_{T}$ range over which the
interfence is visible. }

\textrm{This selection can also be viewed in momentum-space. In the low-energy
limit, the photon flux $n(\omega)$ (Equation \ref{nofomega}) scales as
$1/\omega$. However, when an additional photon is present, the spectrum
becomes much harder. The spectrum for photons that are accompanied by Coulomb
excitation is:
\begin{equation}
n(\omega) \approx\int_{2R_{A}}^{\gamma\hbar c/\omega} d^{2}b N(\omega,b)
\frac{Const.}{b^{2}}\approx Const. \ ;
\end{equation}
The extra photon adds a $1/b^{2}$ weighting, and the resultant flux is
independent of the photon energy. }

\textrm{Figure \ref{join3}(b) compares the spectra with and without tagging.
By selecting reactions with additional accompanying photon interactions,
experimenters can 'tune' their photon beam, hardening or softening the
spectrum. This 'tuning' allows many cross-checks. For example, in vector meson
production, $y= \ln{(2\omega/M_{V}c^{2})}$, but there is a two-fold ambiguity
over which nucleus emitted the photon. By comparing vector meson production
with and without mutual Coulomb excitation, it is possible to account for this
ambiguity and find the production cross section as a function of photon
energy. }

\textrm{The coupling is also very useful in experimental triggering. A simple
reaction like multiple Coulomb excitation can be used to trigger on
small-impact-parameter collisions; these remainder of the event can then be
studied without experimental bias. }

\textrm{For two-photon final states, like $e^{+}e^{-}$ pairs, the situation is
more complex, because the particles are produced outside the nuclei. However,
two-photon reactions are also enhanced at small nucleus-nucleus impact
parameters. The STAR collaboration used mutual Coulomb excitation to study
low-mass $e^{+}e^{-}$ pair production at RHIC, since it was not possible to
trigger on the $e^{+}e^{-}$ pair itself \cite{STAR04}. The presence of the
mutual excitation also significantly hardens the pair mass spectrum. }

\textrm{In multiple photonuclear interactions, the photon polarizations are
collinear. Photons are linearly polarized along the electric field of the
emitting nucleus. In photonuclear interactions, the electric field vector
follows the impact parameter vector. When a single nucleus emits multiple
photons, these photons all have the same linear polarization. When the other
nucleus emits a photon, it will have the opposite polarization. }

\textrm{When the final states are sensitive to the photon polarization, then
angular correlations will be present. In $\rho^{0}$ decay, in the plane
perpendicular to the $\rho^{0}$ direction, the angle between the photon
polarization vector and the $\pi^{+}$ (or $\pi^{-})$ direction is distributed
following a $\cos\theta$ distribution. For the case of two independent
$\rho^{0}$ decays, with uncorrelated photon polarization, the angle between
the two $\pi^{+}$, $\delta\phi$ is evenly distributed between 0 and $2\pi$.
With the polarization correlation, the angular correlation is
\begin{equation}
N(\Delta\phi) \approx1 + \frac{1}{2}\cos(2\Delta\phi).
\end{equation}
A similar distribution is also expected for neutrons from giant dipole
resonance. Correlations between neutron $p_{T}$ in $\rho^{0}$ production and
GDR excitation(s) should also be measurable at RHIC. In the longer term, GDR
excitation neutrons could be used to tag photons according to their
polarization direction, allowing for studies with polarized photons. Similar
polarization should also occur for medium-energy nuclear reactions. }

\textrm{In addition to the correlations due to the geometry, multiple
interactions may be a place to study Bose-Einstein correlations, such as in
$\rho^{0}\rho^{0}$ production by two independent photonuclear interactions.
When the two $\rho^{0}$ are produced on the same nucleus, the production
should be bosonically enhanced (exhibit super-radiance) when the $\rho$ have
(in the nuclear target rest frame) a momentum difference $\Delta p <
\hbar/R_{A}$. }

\section{\textrm{Experimental possibilities and limitations for
ultra-peripheral collisions}}

\textrm{The experimental study of electromagnetic interactions at high energy
colliders is quite new. Since the characteristics of these interactions are
very different from the more common hadronic interactions, most existing and
planned detector systems are not optimized to study them. This section will
discuss some of the general experimental possibilities and limitations at
current and future colliders. }

\textrm{At hadron colliders (in contrast to electron accelerators) it is in
general not possible to detect the outgoing projectiles following an
electromagnetic interaction. This is because of the small momentum transfers
involved. Tagging of nuclei will never be possible because the angular
deflection following the coherent emission of a photon is smaller than the
angular dispersion of the beam. Proton tagging has been proposed, but requires
extremely high resolution Roman pots \cite{Pio01}. }

\textrm{For exclusive particle production, characterized by the emission of
only a few final state particles, good signal to background ratios can be
achieved by selecting events with small total transverse momentum when the
event is reconstructed \cite{KN98}. The total event transverse momentum is the
sum of the momentum transfer from each projectile, which is determined by the
form factors and can be calculated accurately. The method works best for heavy
nuclei. This is illustrated in Figure~\ref{star200_rho0_pt}, which shows the
$p_{T}$ distribution for exclusive $\rho^{0}$ production in Au+Au interactions
at RHIC. The coherent peak for events with exactly one reconstructed $\pi^{+}$
and one reconstructed $\pi^{-}$ can be clearly seen with $p_{T} <$ 100~MeV/c.
The incoherent background can be estimated by measuring events with two
reconstructed pions with the same charge (dashed histogram in the figure). }

\textrm{\begin{figure}[t]
\textrm{\centerline{\psfig{file=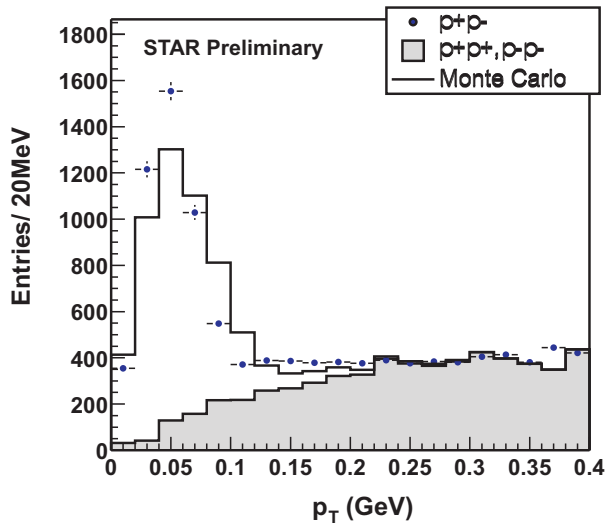,width=8cm}} \vspace*{8pt}
}\caption{ {\small Transverse momentum distribution for events with exactly 2
reconstructed charged pions. Data from the STAR collaboration
\cite{Meissner03}. The Monte Carlo calculation is based on \cite{BKN02}.}}%
\label{star200_rho0_pt}%
\end{figure}}

\textrm{The technique of using the total event $p_{T}$ to identify
electromagnetic interactions works only for coherent and exclusive particle
production. Other photonuclear interactions require different approaches.
Incoherent photon-induced interactions are characterized by a gap in rapidity
void of particles between the rapidity of the photon emitting projectile and
the rapidity of the produced state. This distinguishes the electromagnetic
interactions from ordinary hadronic interactions. }

\textrm{For hadronic interactions, the probability of having a rapidity gap of
width $\Delta y$ void of particles decreases exponentially with increasing
size of the interval:
\begin{equation}
p(n=0) = \exp( - \langle dn/dy \rangle\, \Delta y )
\end{equation}
where $\langle dn/dy \rangle$ is the mean multiplicity per unit of rapidity.
Requiring a rapidity gap of width $\Delta y = 2$, leads to a hadronic
rejection factor of $10^{-2}$ to $10^{-3}$ for single nucleon-nucleon
interactions at RHIC and LHC energies. Additional rejection of hadronic events
can be obtained if the fragmentation of the beam nuclei is detected, for
example in forward (``Zero-degree'') calorimeters. In hadronic interactions
both nuclei normally break up, whereas in photoproduction the photon emitter
normally remains intact. Further discussions of the experimental aspects can
be found in \cite{KNV02}. The conclusion is that sufficient rejection can be
obtained with current and planned detectors at heavy ion colliders to study,
e.g., the production of heavy quark pairs. }

\textrm{One significant background to electromagnetic diffraction is hadronic
diffraction. Pomeron exchange can also produce rapidity gaps, and
double-Pomeron interactions produce isolated systems in a central detector,
with rapidity gaps on each side \cite{FELIX}. These backgrounds are not an
issue for vector meson production, because double-Pomeron interactions do not
produce $J^{PC} = 1^{--}$ final states, but these backgrounds could be
problematic for other reactions. Because the strong force has a short range
($\le1$ fm), Pomeron exchange between nuclei and Double-Pomeron interactions
can only occur between surface nucleons in grazing collisions; this limits the
cross sections. Unfortunately, quantitative estimates are lacking. However,
for pp collisions, Pomeron interactions may constitute a significant
background. }

\textrm{Another experimental challenge is finding an efficient trigger with
adequate background rejection. Since the outgoing beams are not tagged, it is
necessary to trigger on the particles emitted from the final state X to study
the reaction A+A $\rightarrow$ A+A+X, . These particles are usually produced
at or near mid-rapidity. Most of the current and planned heavy-ion experiments
at RHIC and the LHC have triggers that are optimized for hard, high $p_{T}$
interactions. These triggers are difficult to adapt to low multiplicity
low-$p_{T}$ final states. STAR at RHIC is an exception, but it faces
challenges because of the low allowable trigger rates \cite{STARtrigger}.
Backgrounds from beam-gas interactions, grazing hadronic collisions, ambient
neutrons and other beam-related backgrounds are serious concerns to any
experiment. }

\textrm{In combination with the large probability for multiple Coulomb
interactions with heavy beams, factorization makes triggering on neutrons
emitted in the forward direction following single or double Coulomb excitation
an attractive alternative. This reduces the photon flux, but provides some
control of the impact parameter distribution, as discussed in Section 5. }

\textrm{The experimental feasibility of studying ultra-peripheral collisions
is demonstrated through the data presented earlier. However, the
considerations above limit the types of processes that can be studied. The
study of two-photon production of multiple $e^{+} e^{-}$-pairs seems difficult
because of the extremely low $p_{T}$ of the emitted electrons. A dedicated
experiment would do better, but that is likely to be a difficult proposition
at a high-energy physics laboratory. The study of meson spectroscopy at the
LHC also seems problematic because of the lack of triggers for low momentum
charged particles around mid-rapidity. }

\section{\textrm{Conclusions}}

\textrm{We have presented the formalisms for studying photoproduction and
two-photon reactions at hadron colliders, and discussed some of the more
interesting applications of these techniques. Low-energy nuclear physics has
used these techniques for many years, with good results. However, only
recently have higher-energy machines like RHIC and the Tevatron begun to study
particle production in very strong fields. }

\textrm{The amount of experimental data on UPCs is still rather limited.
Despite this, the field has developed tremendously over the past decade, with
much theoretical progress. The new data from RHIC is helping to focus new
theoretical work on the channels that are most readily accessible. With the
new results from RHIC and the coming LHC startup, UPCs are now ready to make
substantial contributions to many areas of physics. }

\textrm{The LHC will produce photonuclear interactions with 10 times the
energy available at other accelerators like HERA. This will open a huge window
to search for new physics processes (some of them not accessible in hadronic
collisions), measure gluon densities at very low Feynman-x, and perform a host
of other measurements. At the same time, UPC reactions are very important for
machine operations - $e^{+} e^{-}$ production with $e^{-}$ capture will limit
the LHC luminosity with heavy ions. }

\textrm{We thank Ramona Vogt for providing Figure \ref{qqbar_rapidity},
Gerhard Baur and Kai Hencken for beneficial discussions, and Heather Gray for
helpful comments on the manuscript. This work was supported by the
U.\thinspace S.\ Department of Energy under grants No. DE-FG02-04ER41338 and
DE-AC-03076SF00098. }

\end{document}